\documentclass[aps,prd,twocolumn,10pt,groupedaddress]{revtex4-1}
\usepackage{amssymb}
\usepackage{graphicx}
\usepackage{amsmath}
\usepackage{hyperref}
\usepackage{color}
\begin{document}

\title{Holographic complexity growth in a FLRW universe}
\author{Yu-Sen An$^{a,b}$}
\email{anyusen@itp.ac.cn}
\author{Rong-Gen Cai$^{a,b}$}
\email{cairg@itp.ac.cn}
\author{Li Li$^{a,b}$}
\email{liliphy@itp.ac.cn}
\author{Yuxuan Peng$^{a,c}$}
\email{yxpeng@itp.ac.cn}

\affiliation{$^a$CAS Key Laboratory of Theoretical Physics, Institute of Theoretical Physics, Chinese Academy of Sciences, Beijing 100190, China}
\affiliation{$^b$School of Physical Sciences, University of Chinese Academy of Sciences, No.19A Yuquan Road, Beijing 100049, P.R. China}
\affiliation{$^c$East China University of Technology, Nanchang, Jiangxi 330013, P.R. China}
\date{\today}

\begin{abstract}
	We investigate the holographic complexity growth rate of a conformal field theory in a FLRW universe. We consider two ways to realize a FLRW spacetime from an  Anti-de Sitter Schwarzschild geometry. The first one is obtained by introducing a new foliation of the Schwarzschild geometry such that the conformal boundary takes the FLRW form. The other one is to consider a brane universe moving in the Schwarzschild background. For each case, we compute the complexity growth rate in a closed universe and a flat universe by using both the complexity-volume and complexity-action dualities. We find that there are two kinds of contributions to the growth rate: one is from the interaction among the degrees of freedom, while the other one from the change of the spatial volume of the universe. The behaviors of the growth rate depend on the details to realize the FLRW universe as well as the holographic conjecture for the complexity. For the realization of the FLRW universe on the asymptotic boundary, the leading divergent term for the complexity growth rate obeys a volume law which is natural from the field theory viewpoint. For the brane universe scenario, the complexity-volume and complexity-action conjectures give different results for the closed universe case. A possible explanation of the inconsistency when the brane crosses the black hole horizon is given based on the Lloyd bound.
	
\end{abstract}

\maketitle

\section{Introduction}
Anti-de Sitter/Conformal field theory (AdS/CFT) correspondence has greatly deepened our understanding of the quantum gravity~\cite{Maldacena:1997re,Gubser:1998bc,Witten:1998qj,Aharony:1999ti}. In particular, motivated by the holographic entanglement entropy~\cite{Ryu:2006bv}, an intrinsic potential connection between quantum information theory and gravity physics has been uncovered. However, in the context of the thermo-field double state (TFD state) which is dual to the eternal black hole~\cite{Maldacena:2001kr}, it has been shown that entanglement entropy can not capture all the information during the evolution of an AdS wormhole~\cite{Hartman:2013qma}. As a more refined information quantity, complexity has been proposed to describe the situation where entanglement entropy fails, such as the wormhole growth behavior far beyond the thermal equilibrium. Both the field theory definition and the holographic definition of complexity have received great attention.  Although there are many investigations on the complexity from field theory side, such as~\cite{Jefferson:2017sdb,Chapman:2017rqy, Yang:2018nda,Yang:2019udi,Yang:2018tpo, Caputa:2017yrh,Bhattacharyya:2018wym,Khan:2018rzm}, a unique and consistent definition is still lacking. From the holographic point of view, there are two proposals for complexity, known as complexity-volume (CV) duality~\cite{Stanford:2014jda} and complexity-action (CA) duality~\cite{Brown:2015bva,Brown:2015lvg}. There are many investigations regarding their properties, such as the growth rate~\cite{Lehner:2016vdi,Miao:2017quj,Carmi:2017jqz,An:2018xhv,Cai:2017sjv,Jiang:2019pgc,Jiang:2019yzs,Mahapatra:2018gig}, the divergence structure~\cite{Carmi:2016wjl,Kim:2017lrw} and the generalization beyond Einstein gravity~\cite{Cai:2016xho,Jiang:2018pfk,Cano:2018aqi,An:2018dbz,Jiang:2019fpz,Jiang:2018sqj}. 

While most works on the complexity growth rate considered the static case, the generalization to the time dependent case is also quite interesting and it is worthwhile to study how the complexity evolves in a dynamical process, for related studies on the Vaidya spacetime, see Refs.~\cite{Chapman:2018dem,Chapman:2018lsv,Jiang:2018tlu}.The investigation of the complexity can also be generalized to states on other dynamical backgrounds which correspond to different slices from gravity side, such as de-Sitter boundary in Ref.~\cite{Reynolds:2017lwq}. Of particular interest is the boundary metric that has the Friedman-Lema\^itre-Robertson-Walker (FLRW) form, which might lead to the understanding of the non-perturbative aspects of cosmology.

In Ref.~\cite{Apostolopoulos:2008ru}, starting with an AdS Schwarzschild black hole, one can choose a different foliation away from the black hole to make the metric time dependent and to realize the boundary with the form of a FLRW spacetime. 
The Friedman equation can also be obtained by considering the mixed boundary conditions on the new slice. Holographically, the dual field theory on the FLRW boundary may represent an expanding plasma and the authors of Ref.~\cite{Apostolopoulos:2008ru} calculated its stress energy tensor and entropy production. 
The paper~\cite{Apostolopoulos:2008ru} adopted the Fefferman-Graham (FG)coordinates. Instead of going to the FG coordinates, the author of Ref.~\cite{Camilo:2016kxq} found a simple foliation of the AdS Schwarzschild black hole and got the same FLRW metric on the boundary. It will be interesting to investigate the complexity behavior of the state on this time dependent boundary, and we hope the results could have some new phenomenon due to the non-equilibrium physics. 

There are also other ways to realize the FLRW cosmology from the bulk AdS-Schwarzschild black hole, such as introducing a co-dimension one brane. The original motivation to consider this realization is from the holographic principle. The relation between cosmology and holography was first raised by Fischler and Susskind~\cite{Fischler:1998st}. After that, Erik Verlinde~\cite{Verlinde:2000wg} investigated the entropy bound and found that the entropy formula called Cardy-Verlinde formula in a CFT can reproduce the Friedman equation, which implies possible connection between CFT and FLRW universe. The various arguments proposed in Ref.~\cite{Verlinde:2000wg} has been naturally realized in the brane-world scenerio in Ref.~\cite{Savonije:2001nd} where the authors embedded the Randall-Sundrum type  \uppercase\expandafter{\romannumeral2} brane in the Schwarzschild AdS black hole.
Randall-Sundrum brane world was first proposed as a solution to the hierarchy problem~\cite{Randall:1999vf,Randall:1999ee}. Maldacena first pointed out that the field theory on the brane should be seen as a CFT coupled to gravity. This idea has been summarized in Refs.~\cite{Gubser:1999vj,Hawking:2000kj}. The spacetime ends on the brane, and the brane can be seen as a time dependent boundary with conformal radiation on it. It is also interesting to investigate the complexity evolution on the brane universe. 

This paper is organized as follows. In Section II, we compute the growth rate of the holographic complexity for the FLRW type boundary theory and show the effect of the time dependence on the complexity growth rate. In Section III, we investigate the complexity growth rate on the brane using both the CV and CA duality conjectures. We consider two cases: a spherical black hole which corresponds to a closed universe and a planar black brane which describes a flat universe. We show the time evolution of the complexity growth rate.  In section IV, we summarize our results and discuss possible future directions. 

\section{Complexity growth on the FLRW type boundary}
This section explores the holographic complexity of some particular FLRW universe which lives on the asymptotic AdS boundary. We briefly introduce the background solution following the setup of Ref.~\cite{Camilo:2016kxq}. Then we study the complexity growth with both the CA and CV conjectures in details.

\subsection{The metric}
The $(d+1)$ dimensional static asymptotically AdS black hole is described by the metric 
\begin{equation}\label{AdSbh}
\mathrm{d}s^{2}=-f(r)\mathrm{d}t^{2}+f(r)^{-1}\mathrm{d}r^{2}+\Sigma(r)^{2} \mathrm{d}\Omega_{k,d-1}^{2}\,,
\end{equation}
where $f(r) \sim r^2/L^2$ and $\Sigma(r) \sim r/L$ at large $r$ with $L$ the AdS radius.
$\mathrm{d}\Omega_{k,d-1}^2$ denotes the line element of the co-dimension two maximally symmetric subspace which can be spherical $(k=+1)$, planar $(k=0)$ or hyperbolic $(k=-1)$, and  we will use $\Omega_{k,d-1}$ to represent the spatial volume of this subspace.
Going to the Eddington-Finkelstein coordinates $\{v, r, \,...\}$ via $dv=dt+dr/f(r)$, we write the metric as
\begin{equation} \label{daf}
\mathrm{d}s^{2}=2 \mathrm{d}v \mathrm{d}r-f(r)\mathrm{d}v^{2}+\Sigma(r)^{2} \mathrm{d}\Omega_{k,d-1}^{2}\,.
\end{equation}
We introduce the new radial coordinate $R=\frac{r}{a(V)}$ and the new time coordinate $V$, $\mathrm{d}v=\mathrm{d}V/a(V)$. Here $a(V)$ is some positive function of $V$. After plugging $\mathrm{d}v$ and $\mathrm{d}r=a(V)\mathrm{d}R+R \dot{a}(V)\mathrm{d}V$ into the metric Eq.(\ref{daf}), and taking the large $R$ limit, one obtains the following time-dependent metric.
\begin{equation}\label{FRWbdy}
\mathrm{d}s^{2}\sim 2\mathrm{d}V \mathrm{d}R +\frac{R^{2}}{L^{2}}[-\mathrm{d}V^{2}+a(V)^{2}\mathrm{d}\Omega_{k,d-1}^{2}]\,.
\end{equation}
It is obvious that the new conformal boundary at $R\rightarrow\infty$ has precisely the desired FLRW form with the time coordinate $V$. Note that such a cosmological boundary is not the same as the commonly used AdS boundary at $r\rightarrow\infty$ where one has a static boundary metric. 
The entropy density is given by the area of the apparent horizon
\begin{equation}
s=\frac{\Sigma(R_{h}a)^{d-1}}{4G}\,,
\end{equation}
where $G$ is the Newton constant and $R_{h}$ is determined by the equation
\begin{equation}
[\partial_{V}\Sigma+(\frac{f(Ra)}{2a^{2}}-R \frac{\dot{a}}{a} )\partial_{R} \Sigma] |_{R=R_{h}}=0\,.
\end{equation}
We can also associate a local temperature to the black hole as
\begin{equation}
T(V)=\frac{T_{H}}{a(V)}\,,
\end{equation}
where $T_{H}$ is the Hawking temperature of the black hole of Eq.(\ref{AdSbh}). In the present paper, we will focus on the Schwarzschild-AdS black hole, for which the blackening factor is
\begin{eqnarray}\label{SchAdS}
f(r)=k+\frac{r^{2}}{L^{2}}-\frac{16\pi G M}{(d-1)\Omega_{k,d-1}r^{d-2}} \,,
\end{eqnarray}
and $\Sigma(r)=r^2/L^2$. Here $M$ is the mass of the black hole.
The energy density can be calculated by using the holographic renormalization procedure, and when $d=4$ the result is
\begin{equation}
\mathcal{E}=\frac{3(\dot{a}^{2}+k)^{2}+12\tilde{M}}{64 \pi G a^{4}}\,,
\end{equation}
with
\begin{eqnarray}
\tilde{M} = \frac{16\pi G M L^2}{3\,\Omega_{k,3}\,r^{2}}\,.
\end{eqnarray}
In the following we will compute the complexity growth rate associated with this FLRW foliation of the Schwarzschild-AdS black hole using both the CV and CA conjectures.
\subsection{ The complexity growth with CV conjecture}
\begin{figure}
	\includegraphics[width=0.55\textwidth]{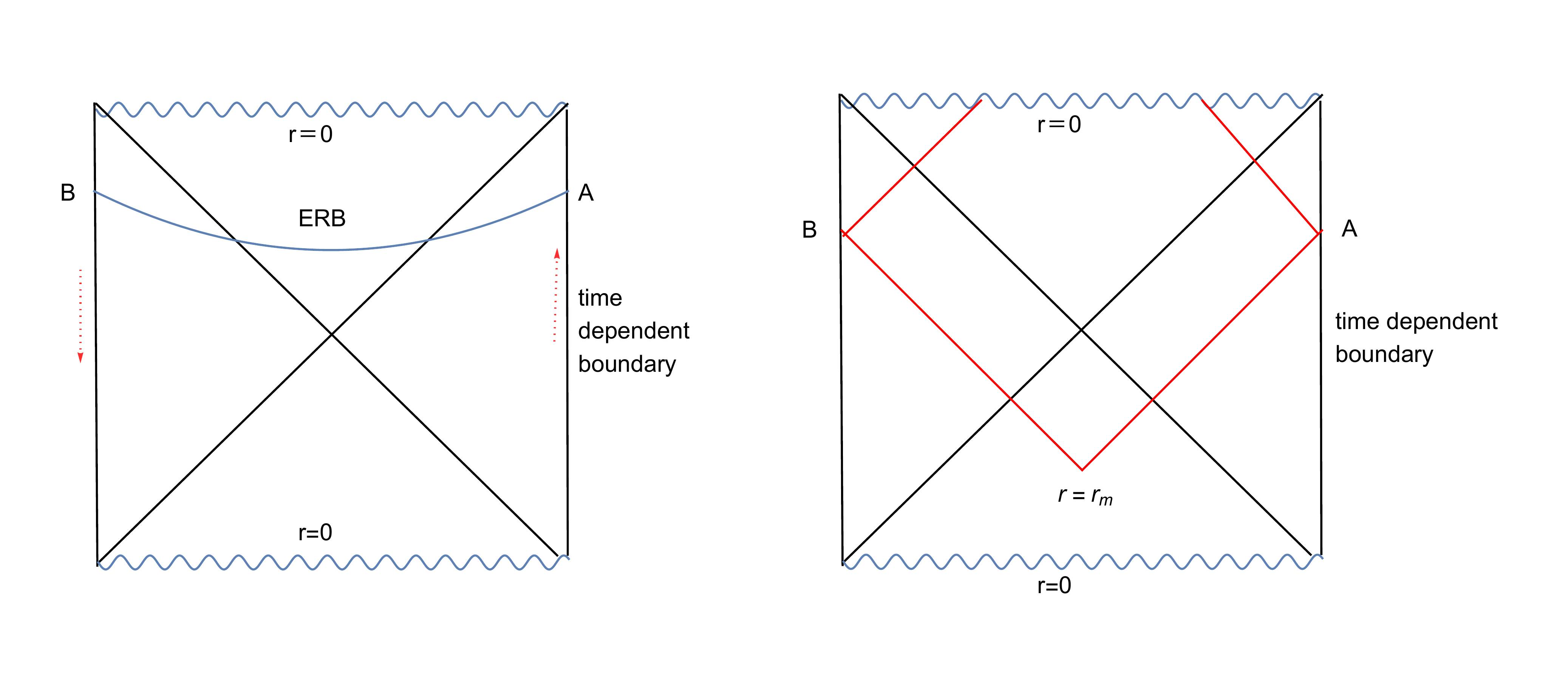}\\
	\caption{Penrose diagram for the two-sided eternal AdS black hole. The CV conjecture is related to the size of an Einstein-Rosen bridge (ERB) to the computational complexity of the dual quantum state (left). The CA conjecture relates the action of the Wheeler-Dewitt patch to the complexity of the CFT state (right).}
	\label{HoloComplexityDiagram}
\end{figure}
The CV and CA methods for computing the holographic complexity are shown schematically in the Penrose diagrams in Fig.~\ref{HoloComplexityDiagram}.
The CV proposal is described in the left panel: the maximal volume of the co-dimension one surface connecting the endpoints on both time-dependent boundaries is proportional to the complexity of the boundary state:
\begin{eqnarray}\label{CV}
C_V=\frac{{\rm max}\left[Volume\right]}{G\ell}\,,
\end{eqnarray}
where $\ell$ is a dimensional constant which is usually chosen to be equal to the AdS radius $L$.
In the original CV proposal the CFT lives on a static boundary, while we will extend the original definition to a time-dependent boundary.
The calculation method mainly follows the procedure provided in Ref.~\cite{Carmi:2017jqz}. Note that the dual state depends on two times $t_L$ and $t_R$ with subscripts $L$ and $R$ representing the left and right boundary times, respectively. We are interested in the symmetric configuration with $t_{L}=t_{R}$. 

As the maximal surface is symmetric with respect to the innermost point of the surface located at the radial coordinate $r_{min}$, we only need to focus on the right side of the maximal surface from $r_{min}$ to an UV cutoff, say at $r_{max}$. Furthermore, the maximal surface has the same symmetry as the horizon, and the volume of the maximal surface can be expressed as 
\begin{eqnarray}
Volume = 2 \Omega_{k,d-1} W\,,
\end{eqnarray} where
\begin{eqnarray}
W= \int_{r_{\rm min}}^{r_{\rm max}} \mathrm{d} \lambda \, \frac{r^{d-1}}{L^{d-1}} \sqrt{-f(r) v'^2 + 2 v'r'}\,,
\end{eqnarray}
with the prime denoting a derivative with respect to $\lambda$ which is the parameter describing the surface. 
Note that the UV cutoff $r_{\rm max}$ will be taken to be infinity finally.
By solving the parameter equations of $v(\lambda)$ and $r(\lambda)$, one can determine the maximal surface and calculate the complexity growth rate. For more details about the computation, one can consult Ref.~\cite{Carmi:2017jqz}.

The endpoint at the right cut-off boundary is covered by the Schwarzschild coordinates $\{t_R, r\}$, and coordinates $\{V_R, R\}$ simultaneously.
The complexity growth rate of our FLRW universe is proportional to the quantity $\partial W/\partial V_R$. According to the chain rule of differentiation, it is given by
\begin{eqnarray}
\frac{\partial W}{\partial V_R}\Big|_{(V_R, R_{\text{max}})} &=&\frac{\partial W(r_{\text{max}}, t_R)}{\partial t_R} \frac{\partial t_R}{\partial V_R}\Big|_{(V_R,R_{\text{max}})} \nonumber\\
&&+ \frac{\partial W(r_{\text{max}}, t_R)}{\partial r_{\text{max}}} \frac{\partial r_{\text{max}}}{\partial V_R}\Big|_{(V_R,R_{\text{max}})}\,.\nonumber\\
\end{eqnarray}
The partial derivatives are given by
\begin{eqnarray}\label{chainrule}
&&\frac{\partial W(r_{\text{max}}, t_R)}{\partial t_R}= - E\,,\nonumber\\ 
&&\frac{\partial t_R}{\partial V_R}= \frac{1}{a(V_R)} - \frac{R_{\text{max}}}{f(r_{\text{max}})}\dot{a}(V_R) \,,\nonumber\\ &&\frac{\partial W(r_{\text{max}}, t_R)}{\partial r_{\text{max}}}=\frac{1}{f(r_{\text{max}})}\sqrt{f(r_{\text{max}})r_{\text{max}}^{2(d-1)}+E^2} \,,\nonumber\\
&&\frac{\partial r_{\text{max}}}{\partial V_R}=\dot{a}(V_R) R_{\text{max}}\,, \nonumber\\
\end{eqnarray}
where the dots denote the derivative with respect to $V_R$, and $r_{\text{max}} = R_{\text{max}} a(V_R)$. We have introduced a quantity $E$ which is the conserved charge on the ERB as $W$ is independent of the coordinate $v$. While $E$ is constant for a given ERB, it changes when the ERB evolves. It approaches $\frac{8 \pi G L M}{(d-1) \Omega_{0,d-1}}$ at the late time limit for $k=0$ case, and has deviations from it due to curvature corrections for $k=1$ and $k=-1$ cases. The quantity $E$ is fixed by
\begin{eqnarray}\label{Econst}
f(r_{\rm min})\, r_{\rm min}^{2(d-1)} +E^2 = 0\,.
\end{eqnarray}
Since the maximal surface is symmetric, the innermost point has the time coordinate $t=0$. 
Moreover, $E<0$ in the upper half (black hole region) of the Penrose diagram.

Taking the limit $R_{\text{max}} \rightarrow \infty$ with $V_R$ fixed, for the Schwarzschild-AdS black hole, Eq.(\ref{SchAdS}), we arrive at
\begin{eqnarray}\label{CVatInfi}
\frac{G L}{2\Omega_{k,d-1}}\frac{\partial C_{V}}{\partial V_R}\Big|_{(V_R, R_{\text{max}})} &\approx&  -\frac{E}{a} -\frac{1}{2}k L^3 a^{d-4}\dot{a} R_{\text{max}}^{d-3} \nonumber \\
&& + \dot{a}  L {R_{\text{max}}^{d-1} a^{d-2}}\nonumber + \cdots\nonumber\\
&=&  -\frac{E}{a} +\frac{L R_{\text{max}}^{d-1}}{d-1}\frac{\mathrm{d}}{\mathrm{d}V_R}(a^{d-1})\nonumber\\
&& -\frac{1}{2}k L^3 a^{d-4}\dot{a} R_{\text{max}}^{d-3}+ \cdots\,.
\end{eqnarray}
where we have used $f(r_{\text{max}})\propto R^2_{\text{max}} a^2/L^2$ at large $R_{\text{max}}$.
Apart from the finite term $-E/a$, the result above contains a leading divergent term (the second term) proportional to the growth rate of the volume of the universe on the boundary. The third term of Eq.(\ref{CVatInfi}) is due to the spatial curvature of the horizon and is vanishing for the planar case, i.e. $k=0$. Other sub-leading divergent terms are denoted by ``$\cdots$''. In particular, there are no such sub-leading divergent terms when $d=4$ as $R_{max} \to \infty$.


\subsection{The complexity growth with CA conjecture}

The CA conjecture is schematically shown in the right panel of Fig.~\ref{AdSbh}.
The red lines are actually null sheets starting from the two endpoints, $A$ and $B$, on the boundary, and  the complexity is proportional to the action in the region surrounded by these sheets, which is called the ``Wheeler-DeWitt(WDW)'' patch:
\begin{eqnarray}
C_A = \frac{\text{Action of WDW patch}}{\pi \hbar}\,,
\end{eqnarray}
with $\hbar$ the reduced Planck constant. The system we will consider is described by the Einstein-Hilbert action with a negative cosmological term, and therefore the blackening factor of Eq.(\ref{AdSbh}) is given by Eq.(\ref{SchAdS}).
The method for the calculation of the action in the presence of null boundary has been developed by Ref.~\cite{Lehner:2016vdi,Parattu:2015gga,Parattu:2016trq},where the action reads 
\begin{equation}
\begin{split}
I = & \frac{1}{16 \pi G} \int_\mathcal{M} d^{d+1} x \sqrt{-g} \left(\mathcal R -2 \Lambda \right) \\
&\quad+ \frac{1}{8\pi G} \int_{\mathcal{B}} d^d x \sqrt{|h|} K + \frac{1}{8\pi G} \int_\Sigma d^{d-1}x \sqrt{\sigma} \eta
\\
&\quad -\frac{1}{8\pi G} \int_{\mathcal{B}'}
d\xi \, d^{d-1} x \sqrt{\gamma} \kappa
+\frac{1}{8\pi G} \int_{\Sigma'} d^{d-1} x \sqrt{\sigma} a +I_{count}\,.
\end{split}
\end{equation}
Terms in the expression above are respectively bulk term, Gibbons-Hawking-York(GHY) boundary term for space-like or time-like boundary, Hayward joint term\cite{Hayward:1993my}, null boundary term, null joint term and counter term needed to cancel the dependence of arbitrary normalization parameter.The joints $\eta$ and $a$ are constructed by the rules summarized in Ref.\cite{Lehner:2016vdi}.$h$,$\sigma$,$\gamma$ is the determinant of the induced metric of the corresponding hyper-surface and $\xi$ is the parameter of null hyper-surface, For simplicity, we will choose affine parametrization and set $ \kappa=0$ in the following, so the contribution of null boundary vanishes. 

Following the analysis in the previous subsection, see in particular Eq.(\ref{chainrule}), the complexity growth rate is given by
\begin{equation}
\begin{split}
\frac{\partial C_{A}}{ \partial V_R} |_{R=R_{max}}=&\left( \frac{1}{a(V_R)} - \frac{R_{\text{max}}}{f(r_{\text{max}})}\dot{a}(V_R) \right)\frac{\partial C_{A}}{\partial t_R}\\&+\dot{a}(V_R) R_{\text{max}} \frac{\partial C_{A}}{\partial r_{max}} \,.
\end{split}
\end{equation}
The term $\frac{\partial C_{A}}{\partial t_R}$ has been already obtained in the literature~\cite{Brown:2015bva,Brown:2015lvg,Carmi:2017jqz}. Note that at the boundary $R_{\rm max} \to \infty$, $f(r) \to({R_{\rm max}a}/{L})^{2}$. So the second term in parentheses vanishes. If we consider the late time limit, the first term will reduce to $\frac{2M}{a(V)}$, where $M$ is the energy of the bulk static spacetime. 

All we need to do is to calculate the second term, i.e. the derivative of the complexity with respect to the $r$ coordinate. In order to do it, we first fix the boundary to be located at a finite position $R=R_{\rm max}$ and then  take the limit $R_{\rm max} \to \infty$. According to Ref.~\cite{Akhavan:2018wla}, the UV cutoff $r=r_{\rm max}$ will also induce a corresponding cutoff surface at $r=r_{0}$ near the singularity at $r=0$, and as $r_{\rm max} \to \infty$, $r_{0}$ goes to $0$. Below we will follow this prescription. We present the formal derivation for general $d$ and take $d=4$ in the final expression. 

The bulk term of the action consists of three parts and we denote the cutoff by $r_{\text{max}}$
\begin{equation}
I_{bulk}^{I}=-\frac{d \Omega_{k,d-1}}{8\pi G L^{2}} \int _{r_{0}}^{r_{h}} r^{d-1}(\frac{t_R}{2}+r^{*}(r_{\text{max}})-r^{*}(r)) \mathrm{d}r\,,
\end{equation}
\begin{equation}
I_{bulk}^{II}=-\frac{d \Omega_{k,d-1}}{8\pi G L^{2}} \int _{r_{h}}^{r_{\text{max}}} r^{d-1} 2 (r^{*}(r_{\text{max}})-r^{*}(r))\mathrm{d}r\,,
\end{equation}
\begin{equation}
I_{bulk}^{III}=-\frac{d \Omega_{k,d-1}}{8\pi G L^{2}} \int _{r_{m}}^{r_{h}} r^{d-1}(-\frac{t_R}{2}+r^{*}(r_{\text{max}})-r^{*}(r)) \mathrm{d}r\,,
\end{equation}
where $r_h$ is the horizon radius, $r^{*}$ denotes the tortoise coordinate defined by $r^{*}=\int \frac{dr}{f(r)}$, and $r_m$ is the radius of the point where the two past null sheets meet with each other, as shown in the right plot of Fig.~\ref{HoloComplexityDiagram}.

The surface term of the cutoff surface inside the horizon is 
\begin{equation}
\begin{split}
I^{f}=- \frac{r^{d-1} \Omega_{k,d-1}}{8\pi G}&(\partial_{r} f(r)+\frac{2(d-1)f(r)}{r})\\& \times (\frac{t_R}{2}+r^{*}(r_{\text{max}})-r^{*}(r)) |_{r=r_{0}}\,.
\end{split}
\end{equation}
There are also various joint terms. 
The joint term at the point $r_{m}$ reads 
\begin{equation}
I_{jnt}=-\frac{\Omega_{k,d-1}r_{m}^{d-1}}{8\pi G} \log \frac{ |f(r_{m})|}{\alpha^{2}}\,,
\end{equation}
where $\alpha$ is a constant normalization parameter of the null normal vector.
The joint term at the surface inside the horizon is 
\begin{equation}
I_{jnt,sing}=-\frac{\Omega_{k,d-1}}{8\pi G} r^{d-1} \log | f(r)|  |_{r=r_{0}}\,.
\end{equation}
Moreover, the two joint terms on the cutoff surfaces are
\begin{equation}
I_{jnt,cut}=\frac{r^{d-1}_{\text{max}} \Omega_{k,d-1}}{4\pi G} \log \frac{f(r_{\text{max}})}{\alpha^{2}}\,.
\end{equation}
In order to eliminate the dependence of the arbitrary choice of the reparameterization, we also add a counter term to the null boundary 
\begin{equation}
I_{count}=-2 \int d^{d-1}x {\rm d}\xi\sqrt{\gamma}\, \Theta \text{log} | \tilde{L}\Theta| \,,
\label{counttt}
\end{equation}
where $\tilde{L}$ is an arbitrary length scale,and $\Theta =({1}/\sqrt{\gamma})( {\partial\sqrt{\gamma}}/{\partial \xi })$ is the expansion. Although this term can modify the full time dependence of the complexity growth, it has no effect on the late time result~\cite{Alishahiha:2018tep}. 

After that, we may take the derivative with respect to $r_{\text{max}}$ and then take $r_{\text{max}}$ to infinity. We see that both the surface term and the joint term inside the horizon are vanishing. So only the bulk term and another three joint terms contribute. 
The first bulk term is
\begin{equation}
\begin{split}
\frac{{\rm d}I_{bulk}^{I}}{{\rm d}r_{\text{max}}}=&\frac{d \Omega_{k,d-1}}{8 \pi G L^{2}} r_{0}^{d-1}(\frac{t_R}{2}+r^{*}(r_{\text{max}})-r^{*}(r_{0}))\frac{{\rm d}r_{0}}{{\rm d}r_{\text{max}}}\\&-\frac{\Omega_{k,d-1}}{8 \pi G L^{2}} \frac{r_{h}^{d}-r_{0}^{d}}{f(r_{\text{max}})}\,.
\end{split}
\end{equation}
Taking the limit $r_{\text{max}} \to \infty$, we find that this term vanishes due to the relation between $r_{0}$ and $r_{max}$~\cite{Akhavan:2018wla}. 
The second bulk term at the boundary reads 
\begin{equation}
\frac{{\rm d}I_{bulk}^{II}}{dr_{\text{max}}}=-\frac{d \Omega_{k,d-1}}{4 \pi G} \int_{r_{h}}^{r_{\text{max}}} \frac{r^{d-1}}{f(r_{\text{max}})}=-\frac{ \Omega_{k,d-1}}{4 \pi G} r_{\text{max}}^{d-2}\,.
\end{equation}
The third bulk term is given by
\begin{equation}
\frac{\mathrm{d}I^{III}_{bulk}}{{\rm d}r_{\text{max}}}=-\frac{d \Omega_{k,d-1}}{8 \pi G L^{2}} \int_{r_{m}}^{r_{h}} \frac{r^{d-1}}{f(r_{\text{max}})} \mathrm{d}r\,,
\end{equation}
where we have used the relation $ -t_R/2+r^{*}(r_{\text{max}})-r^{*}(r_{m})=0$. This term also vanishes by taking $r_{\text{max}} \to \infty$.
 
Next we consider the derivative of the boundary joint terms.
\begin{equation} \label{jntcon}
\begin{split}
\frac{{\rm d}I_{jnt,cut}}{dr_{\text{max}}}=&\frac{(d-1)r_{\text{max}}^{d-2}\Omega_{k,d-1}}{4 \pi G} \log \frac{f(r_{\text{max}})}{\alpha^{2}}\\&+ \frac{r^{d-1}_{\text{max}} \Omega_{k,d-1}}{4 \pi G} \frac{1}{f(r_{\text{max}})} \frac{{\rm d}f(r_{\text{max}})}{{\rm d}r_{\text{max}}}\,.
\end{split}
\end{equation}
The joint term at $r_{m}$ is given by
\begin{equation}
\begin{split}
\frac{{\rm d}I_{jnt}}{dr_{\text{max}}}=\frac{\mathrm{d}I}{dr_{m}} \frac{dr_{m}}{dr_{\text{max}}}=&-\frac{\Omega_{d-1} r_{m}^{d-1}}{8\pi G f(r_{\text{max}})}\frac{{\rm d}f(r_{m})}{{\rm d}r_{m}}\\&-\frac{(d-1)\Omega_{d-1} r_{m}^{d-2}}{8 \pi G} \frac{f(r_{m})}{f(r_{\text{max}})} \log \frac{|f(r_{m})|}{\alpha^{2}}\,,
\end{split}
\end{equation}
where we have used the relation $\frac{dr_{m}}{dr_{\text{max}}}=\frac{f(r_{m})}{f(r_{\text{max}})}$. We find that this term also vanishes by taking the boundary limit $r_{\text{max}} \to \infty$. Now the result depends on the choice $\alpha$, and one needs to consider the counter term which eliminates such arbitrariness. We can take a special parametrization $\xi=r/\alpha$. It is worth noting that the final expression does not depend on the choice of $\alpha$ . Following the result of Ref.~\cite{Carmi:2017jqz}, the counter term contribution is given by
\begin{equation}
\begin{split}
I_{count}=&\frac{\Omega_{k,d-1}}{2\pi G} r_{\text{max}}^{d-1} ( \log \frac{(d-1)\alpha \tilde{L}}{r_{\text{max}}}+\frac{1}{d-1})\\&-\frac{\Omega_{k,d-1}}{4\pi G} r_{m}^{d-1} (\log \frac{(d-1)\alpha \tilde{L}}{r_{m}}+\frac{1}{d-1})\,.
\end{split}
\end{equation}
By taking the derivative with respect to $r_{\text{max}}$, we find that the dependence of $\alpha$ cancels precisely with Eq.(\ref{jntcon}). 
Then we obtain the result
\begin{equation}
\begin{split}
\frac{\partial C_{A}}{\partial r_{\text{max}}} |_{r_{\text{max}}\to \infty}&=\frac{(d-1)\Omega_{k,d-1}}{4\pi G} r_{\text{max}}^{d-2} \log \frac{(d-1)^{2}\tilde{L}^{2} f(r_{\text{max}})}{r_{\text{max}}^{2}} \\&+\frac{r_{\text{max}}^{d-2}\Omega_{k,d-1}}{4\pi G} \\& =(2(d-1)\log\frac{(d-1)\tilde{L}}{L}+1) \frac{r_{\text{max}}^{d-2} \Omega_{k,d-1}}{4\pi G}+ \cdots\,,
\end{split}
\end{equation}
%
where we have used the boundary behavior of $f(r_{\rm max})$ as $r_{\rm max}\rightarrow\infty$, and the sub-leading divergent terms are donoted by ``$\cdots$".
Therefore, we arrival at the final result
\begin{equation} \label{fsfss}
\frac{\partial C_{A}}{\partial V_R} |_{R}=\frac{1}{a} \frac{\partial C_{A}}{\partial t_R}+\frac{\rm d}{\mathrm{d}V_R} (R_{\rm max}^{d-1} a^{d-1}) C_{0}\,,
\end{equation}
with $C_{0}=\frac{2(d-1)\log\frac{(d-1)\tilde{L}}{L}+1}{d-1} \frac{\Omega_{k,d-1}}{4\pi G}$. 

One finds that the complexity growth using the CA conjecture gives very similar behavior as the CV conjecture at leading order of $r_{max}$.  The first term in Eq.(\ref{CVatInfi}) is of just the same form as the first term in Eq.(\ref{fsfss}).
The second terms are the time derivatives of the spatial volume of the universe.
For both conjectures, we will refer to the first term as the ``interaction part'' as it comes from the interaction of the field theory degrees of freedom on the boundary, and the second term as the ``volume part'' since it comes from the change of the spatial volume.
Note that when $a(V)=1$, the FLRW type boundary at $R=\infty$ reduces to the static AdS asymptotic boundary. Eq.(\ref{CVatInfi}) and Eq.(\ref{fsfss}) reduce to$\frac{{\rm d}C_{V}}{{\rm d} t_{R}} =-\frac{2 \Omega_{k,d-1}}{G_{N} L} E$ and $\frac{ {\rm d} C_{A}}{{\rm d}t_{R}}$ respectively, so the evolution is the same as in Ref.~\cite{Carmi:2017jqz}.

\section{complexity growth of the brane cosmology}

\subsection{Brane in the AdS Schwarzschild black hole background}
The above section focuses on the complexity growth of a CFT on a FLRW like background. One disadvantage is that the scale factor $a(\tau)$ can be an arbitrary function and is given by hand.
There is another way to realize the FLRW cosmology from the AdS Schwarzschild black hole, inspired by the Randall-Sundrum model and the holographic principle. The idea is to introduce a lower dimensional brane with a constant tension in the background of the $(d+1)$ dimensional AdS Schwarzschild black hole. The movement of the brane is described by the following boundary action~\cite{Savonije:2001nd}.
\begin{equation}
\mathcal{L}_{b}=\frac{1}{8 \pi G} \int_{\partial M}{\rm d}^d x \sqrt{h} K+ \frac{\kappa}{8 \pi G} \int_{\partial M}{\rm d}^d x \sqrt{h}\,,
\end{equation}
where $K$ is the trace of the extrinsic curvature $K_{ab}$, $\kappa$ is related to the tension of the brane and $h$ is the determinant of the induced metric $h_{ab}$ on the surface of the brane $\partial M$.
By varying this action we obtain the equation of motion of the brane,
\begin{equation}
K_{ab}=\frac{\kappa}{d-1} h_{ab}\,.
\end{equation}

We begin with the bulk geometry
\begin{equation}
{\rm d}s^{2}=\frac{1}{f(a)} \mathrm{d}a^{2}-f(a) \mathrm{d}t^{2}+a^{2} {\rm d}\Omega_{k,d-1}^{2}
\end{equation}
where $f(a)$ is just the function of Eq.(\ref{SchAdS}) with the radius $r$ replaced by $a$. Next, we introduce a new time parameter $\tau$ and take $t$ and $a$ to be $\tau$ dependent with the follwing constraint.
\begin{equation}
\frac{1}{f(a)}(\frac{\mathrm{d}a}{\mathrm{d}\tau})^{2}-f(a) (\frac{\mathrm{d}t}{d \tau})^{2}=-1\,,
\end{equation}
which ensures that $\tau$ is the proper time on the brane.
So the brane is described by the parameter $\tau$ and the $(d-1)$-dimensional cross-section.
On the brane the induced metric takes the form
\begin{equation}
ds_{d}^{2}=-\mathrm{d}\tau^{2}+a(\tau)^{2} d\Omega_{k,d-1}^{2}\,,
\end{equation}
which describes a standard FRW universe with $a(\tau)$ the scale factor.
From the brane equation of motion and the constraint, we can get the relation between $\{t,a\}$ and $\tau$.
\begin{equation}
\frac{\mathrm{d}t}{\mathrm{d}\tau}=\frac{a}{L f(a)}\,,
\label{tandtau}
\end{equation}
\begin{equation}\label{aandtau}
\left(\frac{\mathrm{d}a}{\mathrm{d}\tau}\right)^2= \frac{a^2}{L^2} - f(a)\,.
\end{equation}
where we have set $\kappa=1/L$ as Ref.~\cite{Savonije:2001nd}.

In this case, we can see that the scale factor on the brane can be deduced from the equation of motion, Eq.(\ref{aandtau}), once the background is fixed. It has been argued by Maldacena that the brane world should be interpreted as a CFT on the brane coupled to gravity. As the conformal field is coupled to gravity, the behavior of complexity will be more complicated than the previous case on the asymptotic AdS boundary. In contrast to the previous section, the brane is now located at finite radius. So the result is free from divergence. 

By considering the time coordinate on the right brane, the growth rate can be obtained by the chain rule
\begin{eqnarray}\label{chainrulebrane}
\frac{{\rm d}C}{{\rm d}\tau} = \frac{{ \partial}C}{{\partial}t_R}\frac{{\rm d}t_R}{{\rm d}\tau} + \frac{{ \partial}C}{{\partial}a}\frac{{\rm d}a}{{\rm d}\tau}\,.
\end{eqnarray}
There are two different effects in the equation above.
The first term is just the same as the first term in Eq.(\ref{CVatInfi}) or Eq.(\ref{fsfss}).
So we also call it the ``interaction part''.
The second term includes the contribution of the volume change, just like the second term in Eq.(\ref{CVatInfi}) or Eq.(\ref{fsfss}), so we also refer to it as the ``volume part''.

\begin{figure}
	\includegraphics[width=0.55\textwidth]{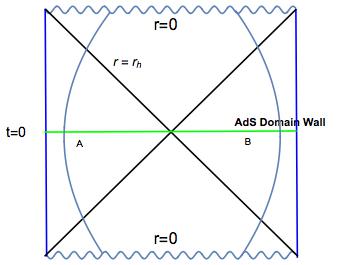}\\
	\caption{AdS domain wall for a spherical Schwartzchild black hole which corresponds to a closed universe that expands to a certain size and then contracts.}	\label{AdSdomainwall1}
\end{figure}

\begin{figure}
	\includegraphics[width=0.55\textwidth]{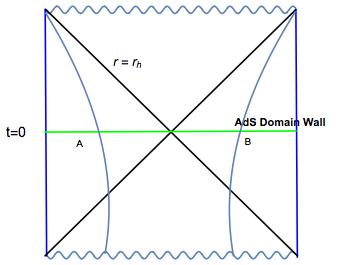}\\
	\caption{AdS domain wall for a planar Schwartzchild black brane which is related to a flat universe that will expand forever.}	\label{AdSdomainwall2}
\end{figure}

In the following we still stick to the picture of symmetric objects (maximal surface/WDW action) for both CV and CA proposals, and obtain the final results by considering the right side of the object and the endpoint $(t_R, a)$ on the right side brane.
We first examine the $k=1$ case corresponding to a closed universe that first expands and then contracts on the brane (see Fig.~\ref{AdSdomainwall1}).
Then we consider the $k=0$ case which corresponds to an ever-expanding open universe on the brane, shown in Fig.~\ref{AdSdomainwall2}.

\subsection{Complexity evolution for the closed universe: the CV conjecture}

According to the chain rule Eq.(\ref{chainrulebrane}), the growth rate is given by
\begin{eqnarray}\label{chainrule2}
	\frac{\mathrm{d} C_V}{\mathrm{d} \tau} &=& \frac{2\Omega_{k,d-1}}{GL}\left(- E \frac{{\rm{d}} t_R}{\rm{d} \tau} + \frac{1}{f(a)}\sqrt{f(a)a^{2(d-1)}+E^2} \frac{{\rm{d}} a}{\rm{d} \tau}\right)\,.\nonumber\\
\end{eqnarray}
On the endpoint of the maximal surface, the time coordinate $t_R$ is expressed by (see Ref.~\cite{An:2018dbz} for more details)
\begin{eqnarray}\label{tauofrmin}
	t_R(\tau) &=&\int^{a(\tau)}_{r_{\rm min}} \mathrm{d}r \frac{E}{f(r)\sqrt{f(r) r^{2(d-1)} +E^2}}\,,\nonumber\\
	&=& -\int^{a(\tau)}_{r_{\rm min}} \mathrm{d}r \frac{\sqrt{-f(r_{\rm min})}\,  r_{\rm min}^{d-1}}{f(r)\sqrt{f(r) r^{2(d-1)} - f(r_{\rm min})\, r_{\rm min}^{2(d-1)}}}\,.\nonumber\\
\end{eqnarray}
On the other hand, $t_R(\tau)$ should also satisfy the diffrential equation Eq.(\ref{tandtau}).
Therefore, one can find out the relation between $r_{\rm min}$ and $\tau$ by combining Eqs.(\ref{tandtau}) and (\ref{tauofrmin}).
Then the complexity growth rate can be obtained by putting $r_{\rm min}$ and $\tau$ into Eqs.(\ref{Econst}) and (\ref{chainrule2}).

First, we should determine the location and shape of the brane. As we can see from Eqs.(\ref{tandtau}) and (\ref{aandtau}), they depend on the spatial curvature, spacetime dimension as well as the location of the horizon.
In the present section we will consider the closed universe in 4 dimensions, so we fix $k=1$ and $d=4$. As we will show that the complexity growth rate exhibits distinct behaviors for small and large values of the horizon radius. 

\begin{figure}
	\includegraphics[width=0.5\textwidth]{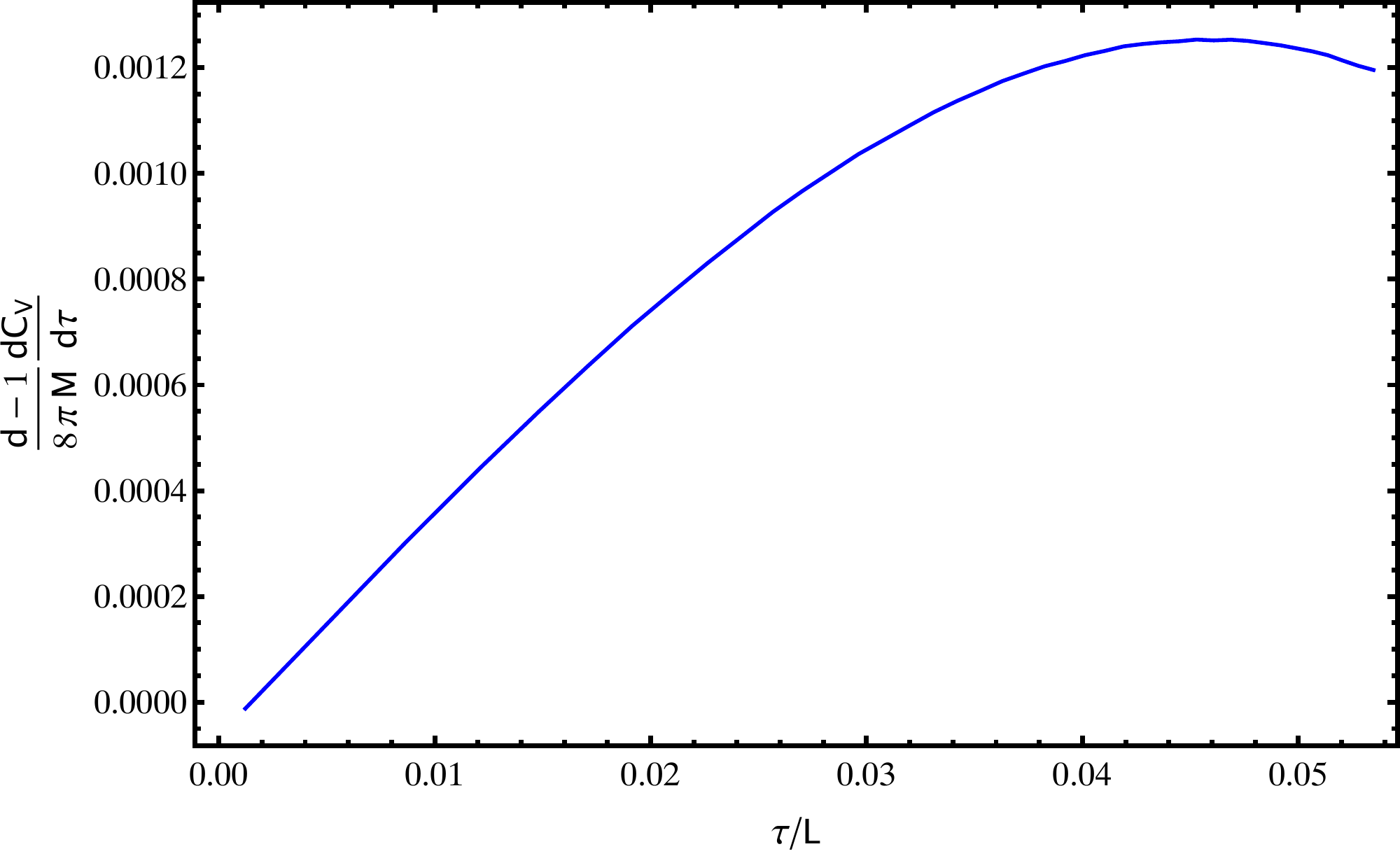}\\
	\caption{The complexity growth rate of the brane cosmology for the closed universe in the CV conjecture. The dimensionless quantities are $\tau/L$ and $(d-1)/(8\pi M)\mathrm{d}C_V/\mathrm{d}\tau$. Here we have chosen $d=4$ and $r_h/L=1$. This figure shows the range $0<\tau/L<0.054$.}\label{CVk11}
\end{figure}

\begin{figure}
	\includegraphics[width=0.5\textwidth]{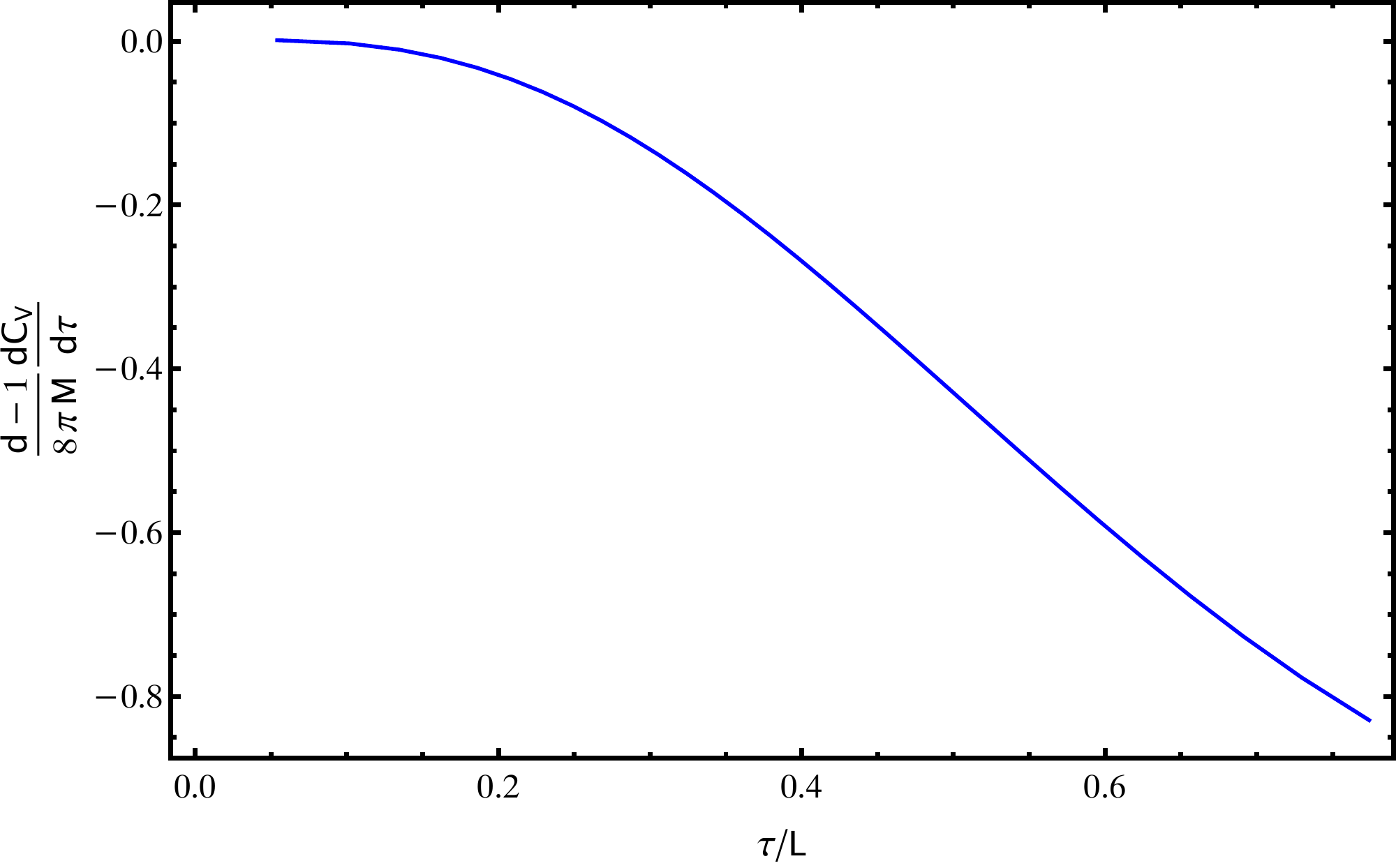}\\
	\caption{The complexity growth rate of the brane cosmology for the closed universe in the CV conjecture. The dimensionless quantities are $\tau/L$ and $(d-1)/(8\pi M)\mathrm{d}C_V/\mathrm{d}\tau$. Here we have considered the case with $r_h/L=1$ and $d=4$. This figure shows the range $0.054<\tau/L<1$.}\label{CVk12}
\end{figure}

As a concrete example, let's first consider the case with the horizon radius at $r_h=L$. By integrating Eq.(\ref{aandtau}) we obtain $\frac{a(\tau)}{L}=\sqrt{2-\frac{\tau^{2}}{L^{2}}}$ with $\tau$ from $-\sqrt{2} L$ to $\sqrt{2} L$, which corresponds to the universe that first expands and then contracts. The relation between $t$ and $\tau$ is
\begin{equation}
t=\int_{-\sqrt{2}L}^{\tau} \frac{a \mathrm{d}\tau}{L f(a)}+t_i\,,
\label{tanddtau}
\end{equation}
with $t_i$ the integration constant. As the evolution of the brane is symmetric, it will be convenient to choose $t_i$ by the condition that $t=0$ when $\tau=0$.We are interested in the contraction phase of this closed universe starting from $\tau=0$ with the initial state being $| TFD(\tau=0) \rangle$. Comparing with Ref.~\cite{Carmi:2017jqz}, here we consider the evolution of TFD state in the contracting FLRW background. When $\tau/L>1$, the brane begins to cross the horizon of the AdS black hole and our description would become un-trustable due to quantum corrections~\cite{Verlinde:2000wg}. So in the following discussion, we will restrict ourselves to the time range before the brane crosses the horizon. We shall return to this point later.
The time evolution behavior of the complexity growth rate is shown in Fig.~\ref{CVk11} and Fig.~\ref{CVk12}.

\begin{figure}
	\includegraphics[width=0.5\textwidth]{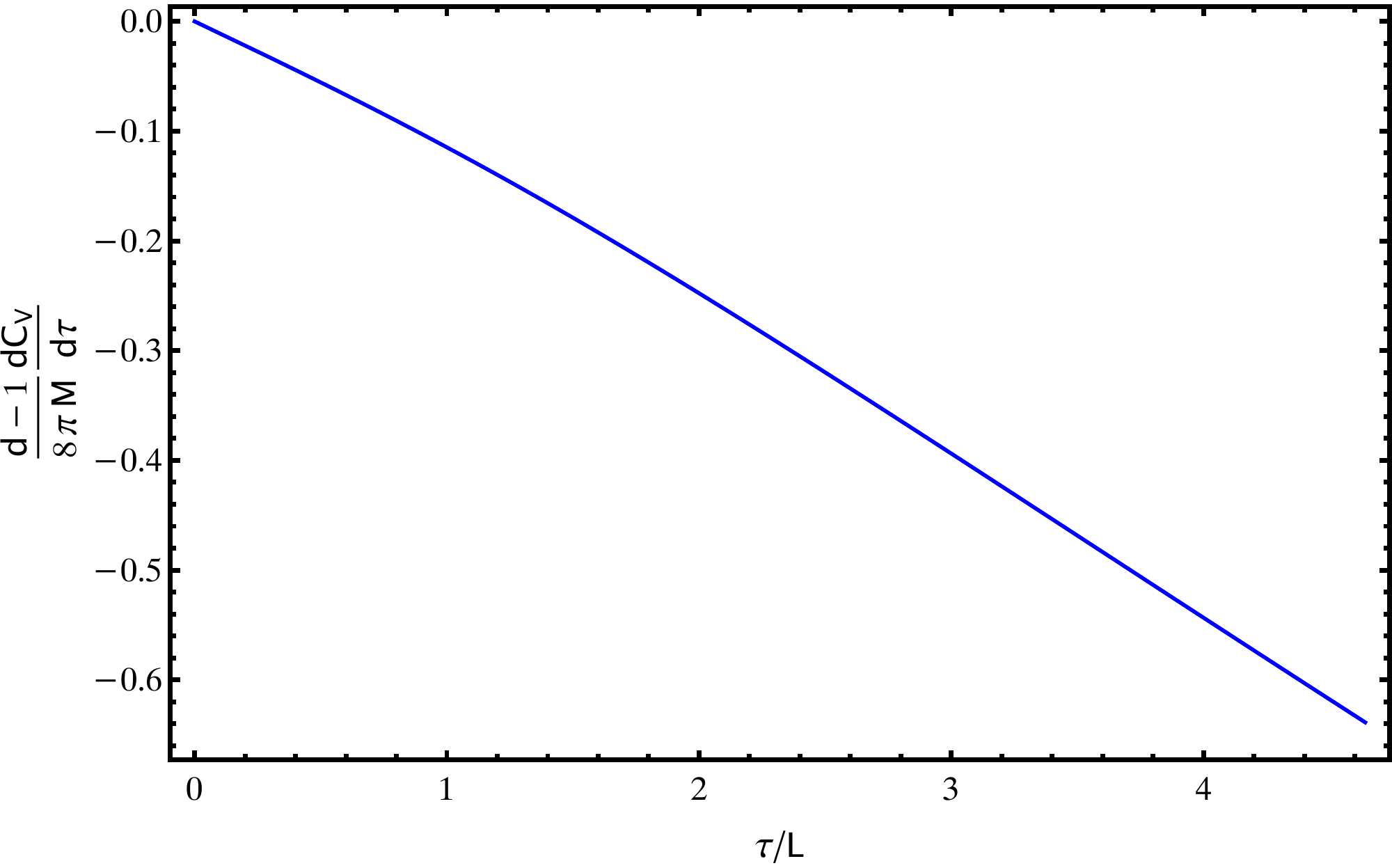}\\
	\caption{The complexity growth rate of the brane cosmology for the closed universe in the CV conjecture. We choose a larger value of the horizon radius with $r_h/L=5$ and dimension $d=4$. The dimensionless quantities are $\tau/L$ and $(d-1)/(8\pi M)\mathrm{d}C_V/\mathrm{d}\tau$. This figure shows the range $0<\tau/L< 5$.}\label{CVk13}
\end{figure}

The non-monotonic behavior of the complexity growth is observed in this case. The growth rate first rises as the time evolves, arrives at its maximum at a certain time, and then it decreases monotonously. It becomes negative at late time. One finds that the complexity will first increases and then decreases, even though the universe is in a contracting phase.
However, if one increases the radius of the black hole, there will be no such non-monotonic behavior. As one can see from Fig.~\ref{CVk13}, the complexity growth rate is negative and decreases all the time from $\tau=0$. So the complexity of such contracting universe decreases faster and faster.

\subsection{Complexity evolution for the closed universe: the CA conjecture}
In this section we turn to the complexity growth rate for the closed universe by using the CA conjecture. The structure of the WDW patch is time dependent. There is a critical time at $\tau_{c}$($t_c$) before which the WDW patch intersects two singularities and above which an additional joint term forms due to the intersection of the past two null segments. The critical time is can be obtained by the relation $t_{c}=\int_{0}^{a_{c}} \frac{\mathrm{d}a}{f(a)}$,where $a_{c}$ is the corresponding position of the brane at that critical time.



Before the critical time, the action contains the bulk term, the past and future surface terms, and two joint terms at A and B (see Fig.~\ref{AdSdomainwall1}). The bulk part consists of three portions.
\begin{equation}
I^{1}_{bulk}=-\frac{d \Omega_{1,d-1}}{4 \pi G L^{2}} \int_{0}^{r_{h}} r^{d-1}(t+r^{*}(a)-r^{*}(r))\mathrm{d}r\,,
\end{equation}
\begin{equation}
I_{bulk}^{2}=-\frac{d \Omega_{1,d-1}}{2 \pi G L^{2}} \int_{r_h}^{a} r^{d-1}(r^{*}(a)-r^{*}(r)) \mathrm{d}r\,,
\end{equation}
\begin{equation}
I^{3}_{bulk}=-\frac{d \Omega_{1,d-1}}{4 \pi G L^{2}} \int_{0}^{r_{h}} r^{d-1}(-t+r^{*}(a)-r^{*}(r))\mathrm{d}r\,.
\end{equation}
Note that here we denote $t$ to be the time on one side boundary, and the total time is $2t$.
The surface term is given by
\begin{equation}
\begin{split}
I_{surf}^{past}=-\frac{r^{d-1} \Omega_{1,d-1}}{8 \pi G}&(\partial_{r} f(r)+\frac{2(d-1)}{r}f(r))\\& \times (-t+r^{*}(a)-r^{*}(r)) |_{r=\epsilon}\,,
\end{split}
\end{equation}
\begin{equation}
\begin{split}
I_{surf}^{future}=-\frac{r^{d-1} \Omega_{1,d-1}}{8 \pi G}&(\partial_{r} f(r)+\frac{2(d-1)}{r}f(r))\\& \times (t+r^{*}(a)-r^{*}(r)) |_{r=\epsilon}\,,
\end{split}
\end{equation}
where $\epsilon$ is the infinitesimal cutoff near $r=0$. Finally, the joint term at the brane reads
\begin{equation}
I_{jnt}^{A+B}=\frac{a^{d-1} \Omega_{1,d-1}}{4\pi G} \log \frac{|f(a)|}{\alpha^{2}}\,,
\end{equation}
where $\alpha$ is the normalization constant. As one can see that the joint term depends on the affine parameter whose choice is quite general. To cancel this ambiguity, one needs to add the counter terms given by Eq.(\ref{counttt}), for simplicity we take $\tilde{L}=L$.
%
%
For the time before $t_{c}$,
\begin{equation}
I_{count}=\frac{\Omega_{1,d-1}}{2\pi G} a^{d-1} (\log \frac{(d-1)\alpha L}{a}+\frac{1}{d-1})\,.
\end{equation}

\begin{figure}
	\includegraphics[width=0.5\textwidth]{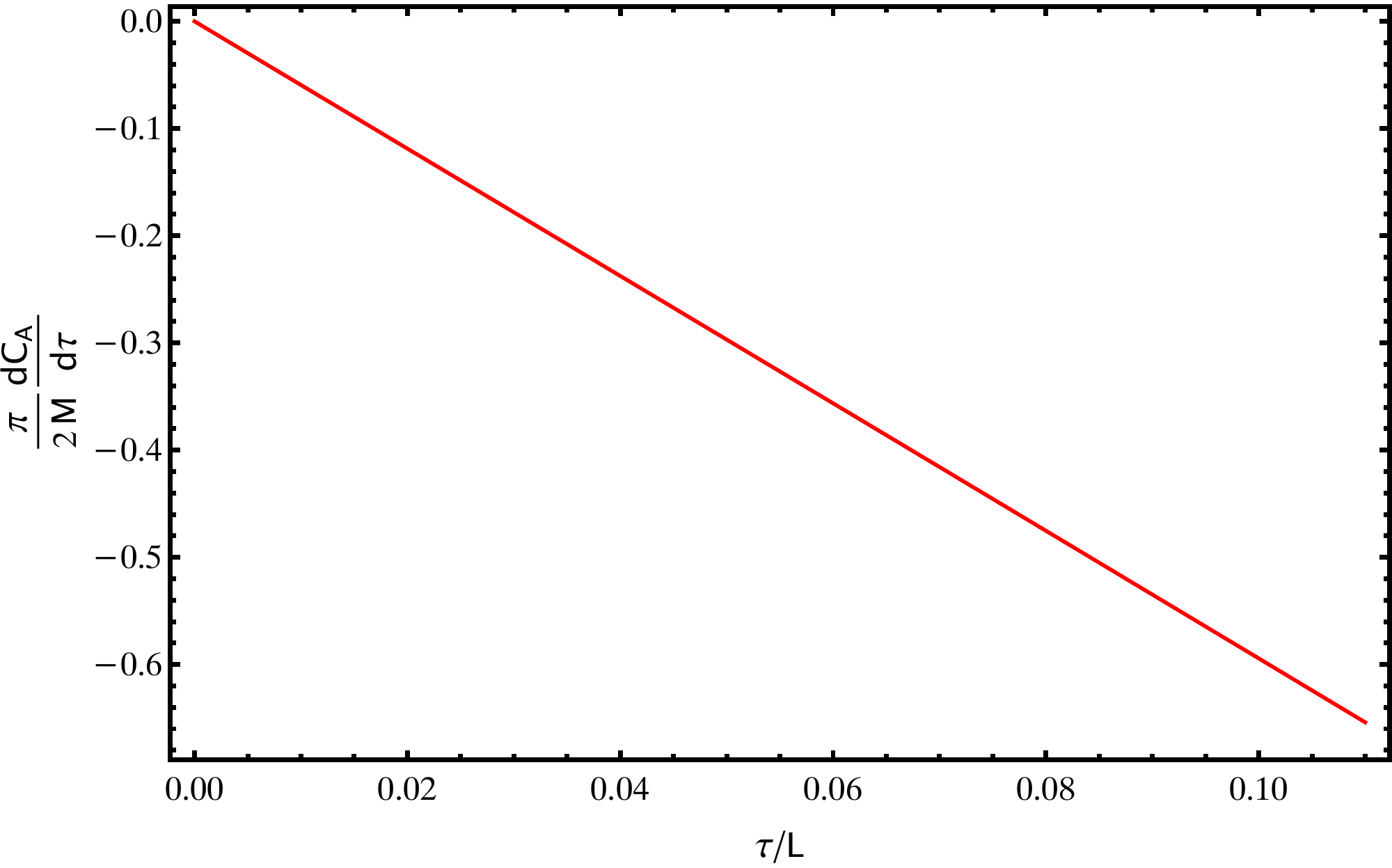}\\
	\includegraphics[width=0.5\textwidth]{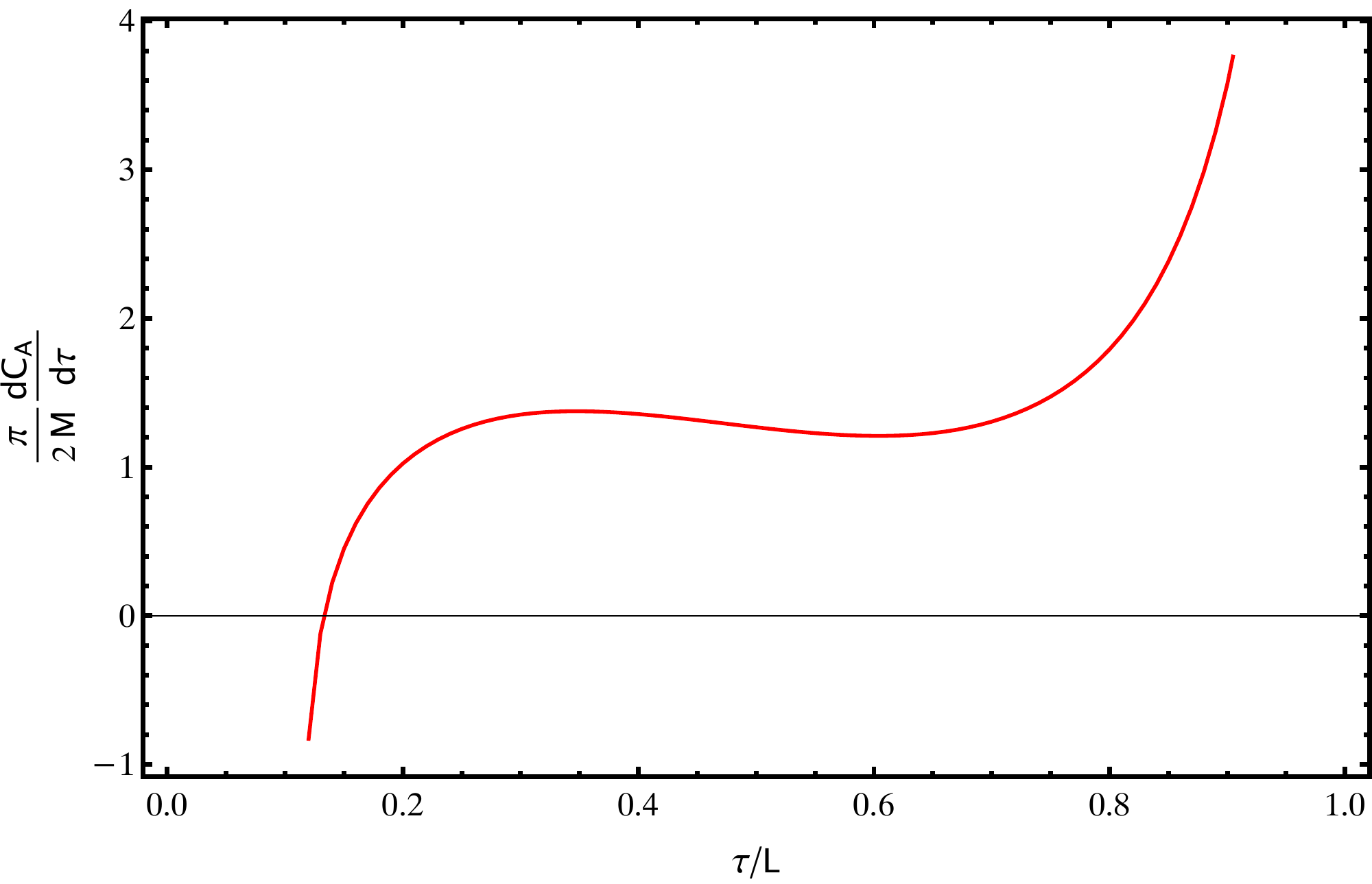}
	\caption{Time dependence of complexity for a closed universe on a brane.
		We have adopted the CA conjecture and set $r_h/L=1$ with the critical time $\tau_{c}/L=0.11$.
		Top panel: the growth rate of the complexity before the critical time $\tau_{c}$, Bottom panel: the growth rate after the critical time.
	}
	\label{CAgrowth}
\end{figure}

Now we are ready to calculate the complexity growth rate. The bulk contribution is 
\begin{equation}
\frac{d I_{bulk}}{\mathrm{d}\tau}=-\frac{\Omega_{1,d-1}}{2 \pi G L^{2}} \frac{a^{d}}{f(a)} \frac{\mathrm{d}a}{\mathrm{d}\tau}\,.
\end{equation}
The contribution from the two surface terms is 
\begin{equation} \label{cf}
\frac{{\rm d}I_{surf}^{F}}{\mathrm{d}\tau}=\frac{d r_{h}^{d-2} \Omega_{1,d-1}}{8 \pi G} (1+\frac{r_{h}^{2}}{L^{2}})( \frac{\mathrm{d}t}{\mathrm{d}\tau}+\frac{1}{f(a)} \frac{\mathrm{d}a}{\mathrm{d}\tau})\,,
\end{equation}
\begin{equation}
\frac{{\rm d}I_{surf}^{P}}{\mathrm{d}\tau}=\frac{d r_{h}^{d-2} \Omega_{1,d-1}}{8 \pi G} (1+\frac{r_{h}^{2}}{L^{2}})(- \frac{\mathrm{d}t}{\mathrm{d}\tau}+\frac{1}{f(a)} \frac{\mathrm{d}a}{\mathrm{d}\tau})\,.
\end{equation}
The part from the joint term at the brane reads
\begin{equation}
\begin{split}
\frac{{\rm d}I_{jnt}(a)}{d \tau}= &\frac{(d-1) \Omega_{1,d-1}a^{d-2}}{4\pi G} \log \frac{f(a)}{\alpha^{2}} \frac{\mathrm{d}a}{\mathrm{d}\tau}\\&+\frac{a^{d-1} \Omega_{k,d-1}}{4\pi G} \frac{1}{f(a)} \frac{{\rm d}f(a)}{\mathrm{d}a} \frac{\mathrm{d}a}{\mathrm{d}\tau}\,.
\end{split}
\end{equation}
The counter term contribution is 
\begin{equation}
\frac{{\rm d}I_{count}}{\mathrm{d}\tau}=\frac{(d-1)\Omega_{1,d-1}}{2\pi G} a^{d-2}\log \frac{\alpha L (d-1)}{a} \frac{\mathrm{d}a}{\mathrm{d}\tau}\,,
\end{equation}
and the contribution from the joint term at the singularity vanishes. One can see that the whole time dependence of the complexity comes from $a(\tau)$. We show the complexity growth rate with respect to the time $\tau$ before $\tau_c$ in the top panel of Fig.~\ref{CAgrowth}. The rate is negative and decreases as $\tau$ is increased from $\tau=0$ to $\tau=\tau_c$.

After the critical time but before crossing the horizon, the action has an additional joint term which is due to the intersection of two past null surface, say at $r_{m}$. 
Such joint term contribution is given by
\begin{equation}
I_{jnt}=-\frac{\Omega_{d-1} r_{m}^{d-1}}{8 \pi G} \log \frac{|f(r_m)|}{\alpha^{2}}\,,
\end{equation}
with $r_{m}$ determined by
\begin{equation} \label{rmm}
t+r^{*}(r_{m})-r^{*}(a)=0\,.
\end{equation}
So the growth rate of this joint term is 
\begin{equation}
\begin{split}
\frac{{\rm d}I_{jnt}(r_{m})}{d \tau}=&- \frac{(d-1) \Omega_{1,d-1}r_{m}^{d-2}}{8\pi G} \log \frac{|f(r_{m})|}{\alpha^{2}} \frac{dr_{m}}{\mathrm{d}\tau}\\&-\frac{r_{m}^{d-1} \Omega_{k,d-1}}{8\pi G} \frac{1}{f(r_{m})} \frac{{\rm d}f(r_{m})}{dr_{m}} \frac{dr_{m}}{\mathrm{d}\tau}\,.
\end{split}
\end{equation}
The bulk parts read
\begin{equation}
I^{1}_{bulk}=-\frac{d \Omega_{1,d-1}}{4 \pi G L^{2}} \int_{0}^{r_{h}} r^{d-1}(t+r^{*}(a)-r^{*}(r))\mathrm{d}r\,,
\end{equation}
\begin{equation}
I^{2}_{bulk}=-\frac{d \Omega_{1,d-1}}{2 \pi G L^{2}} \int_{r_{h}}^{a} r^{d-1}(r^{*}(a)-r^{*}(r))\mathrm{d}r\,,
\end{equation}
\begin{equation}
I^{3}_{bulk}=-\frac{d \Omega_{1,d-1}}{4 \pi G L^{2}} \int_{r_{m}}^{r_{h}} r^{d-1}(-t+r^{*}(a)-r^{*}(r))\mathrm{d}r\,,
\end{equation}
and therefore,
\begin{equation}
\frac{{\rm d}I_{bulk}}{\mathrm{d}\tau}=-\frac{\Omega_{1,d-1}r_{m}^{d}}{4\pi G L^{2}} \frac{\mathrm{d}t}{\mathrm{d}\tau}+\frac{\Omega_{1,d-1}}{4\pi G L^{2}}\frac{r_{m}^{d}-2a^{d}}{f(a)} \frac{\mathrm{d}a}{\mathrm{d}\tau}\,.
\end{equation}
Now there is only one future boundary term, Eq.(\ref{cf}). 
The joint term at the brane does not change, but the counter term and its growth rate become different. 
\begin{equation}
\begin{split}
I_{count}=&\frac{\Omega_{1,d-1}}{2\pi G} a^{d-1} (\log \frac{(d-1)\alpha L}{a}+\frac{1}{d-1})\\&-\frac{\Omega_{k,d-1}}{4 \pi G} r_{m}^{d-1}(\log \frac{(d-1) \alpha L}{r_{m}}+\frac{1}{d-1})\,,
\end{split}
\end{equation}
\begin{equation}
\begin{split}
\frac{{\rm d}I_{count}}{\mathrm{d}\tau}=&\frac{(d-1)\Omega_{1,d-1}}{2\pi G} a^{d-2}\log \frac{\alpha L (d-1)}{a} \frac{\mathrm{d}a}{\mathrm{d}\tau}\\&-\frac{(d-1)\Omega_{k,d-1}}{4\pi G} r_{m}^{d-2}\log \frac{\alpha L (d-1)}{r_{m}} \frac{dr_{m}}{\mathrm{d}\tau}\,,
\end{split}
\end{equation}
where $dr_{m}/\mathrm{d}\tau$ can be obtained from Eq.(\ref{rmm}).
Combining them together, we obtain the complexity growth rate.

We show the time evolution of the growth rate in Fig.~\ref{CAgrowth}. As one can see, although the volume of the universe decreases, the growth rate is first negative but then suddenly becomes positive after the critical time $\tau_{c}$. Here we have chosen $r_h=L$, but we have similar behaviors for other values of $r_h$.
Note that the contribution from the ``interaction part" and the ``volume part" to the complexity growth has opposite effects. The ``interaction part" contribution may become dominant and the complexity continues to grow even if the volume is contracting. In particular, when the brane moves close to the horizon (after $\tau \approx 0.8$ in Fig.~\ref{CAgrowth}), the complexity growth rate increases very quickly and tends to diverge.
Such unnatural behavior motivates us to conjecture that there might be some inconsistency when the brane is very close to the horizon. We will show some evidence based on the Lloyd bound~\cite{Lloyd:2000} in the discussion section.

\subsection{Complexity evolution for the spatially flat universe}

For the AdS spacetime, apart from the black hole with  spherical topology, there are also black hole solutions with flat or hyperbolic horizons. 
It turns out that, for the $k=0$ black brane geometry, the co-dimension one brane embedded in this background represents an expanding flat universe and some of its thermodynamic behavior was discussed in Ref.~\cite{Youm:2001yq}. In this section we want to investigate the complexity behavior on this flat universe. Most of the steps for the calculation will be the same as the closed universe case, so we will skim over the details and will show the main results only.

First, we need to determine the evolution of the brane universe. We focus on the four dimensional universe with $d=4$, for which the blackening factor reads $f(a)=\frac{a^2}{L^2}(1-r_h^4/ a^4)
$ with $r_h$ the location of the horizon.
The differential equation Eq.(\ref{aandtau}) now becomes
\begin{equation}
\frac{\mathrm{d}a(\tau)}{{\rm d} \tau}= \frac{r_{h}^{2}/L}{a(\tau)}\,,
\end{equation}
and the relation between the time $t$ and $\tau$ is also given by Eq.(\ref{tandtau}). As a typical example, we set $r_h/L=1$, then the scale factor is given by $a(\tau)=\sqrt{2 \tau L}$. So the universe will expand forever and never contract. For the present case, the time when the brane universe crosses the horizon is at $\tau=\frac{1}{2} L$. There is a free parameter 
$t_{0}$ when the brane crosses the horizon by solving Eq~(\ref{tandtau}) , this parameter just labels how far the two flat universes are from each other. Below,we choose a specific case with $t_{0}=-6L$. 

\begin{figure}
	\includegraphics[width=0.5\textwidth]{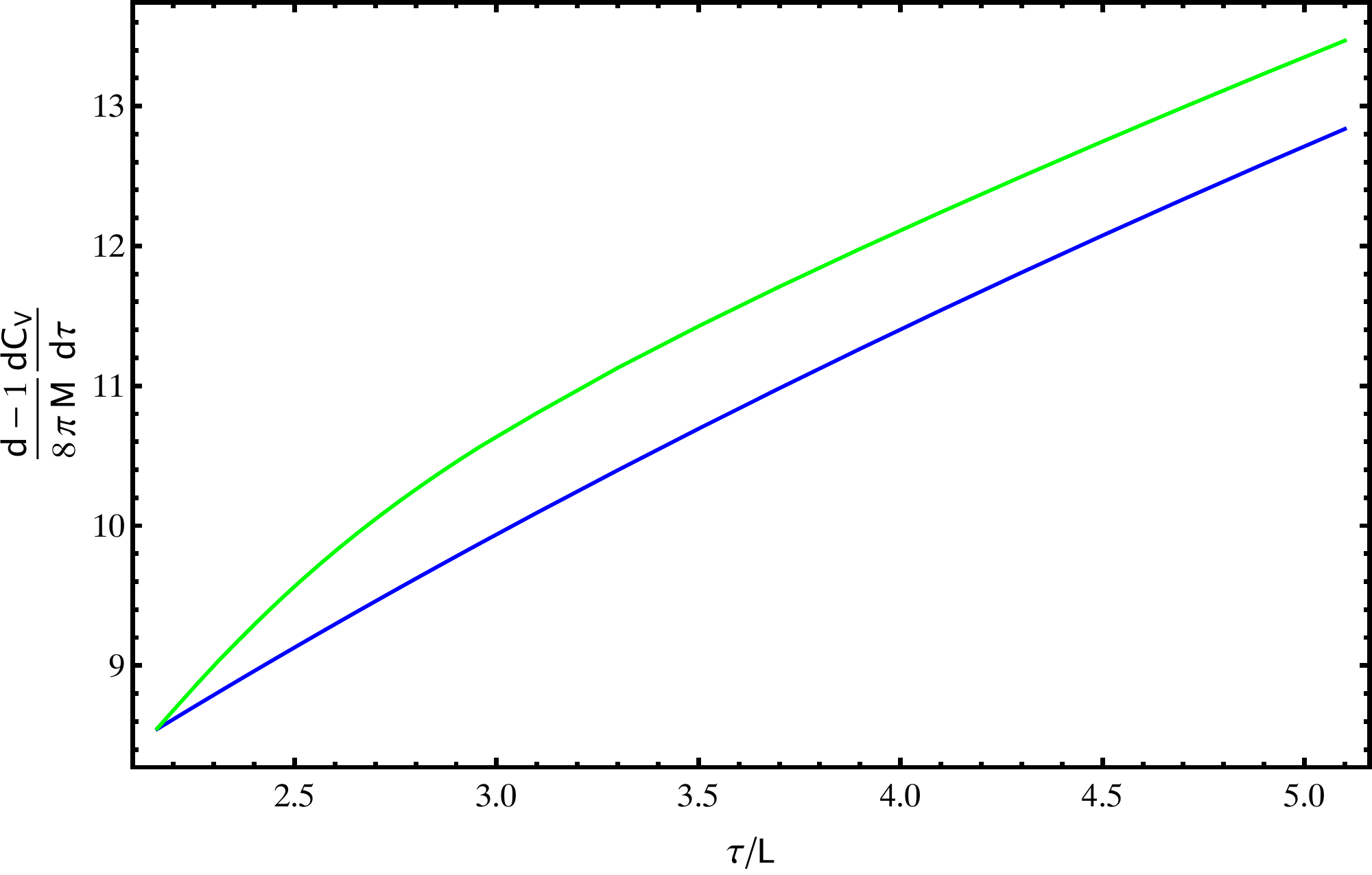}\\
		\caption{The complexity growth rate of the CV proposal for a $d=4$ flat universe in the background of a black brane. The figure starts from $t=0$ with $\tau\approx 2.16L$. The green curve is the total complexity growth rate, and the blue curve is the one due to the volume expansion. We see that the contribution comes mainly from the volume expansion.}\label{CVk01}
\end{figure}

The time-evolution behavior from the CV proposal is shown in Fig.~\ref{CVk01}.
One can see that the complexity growth rate is positive and increases all the time. It is quite different from the case for the closed universe in Fig.4, where the complexity growth rate first increases and then decreases. 

\begin{figure}
	\includegraphics[width=0.5\textwidth]{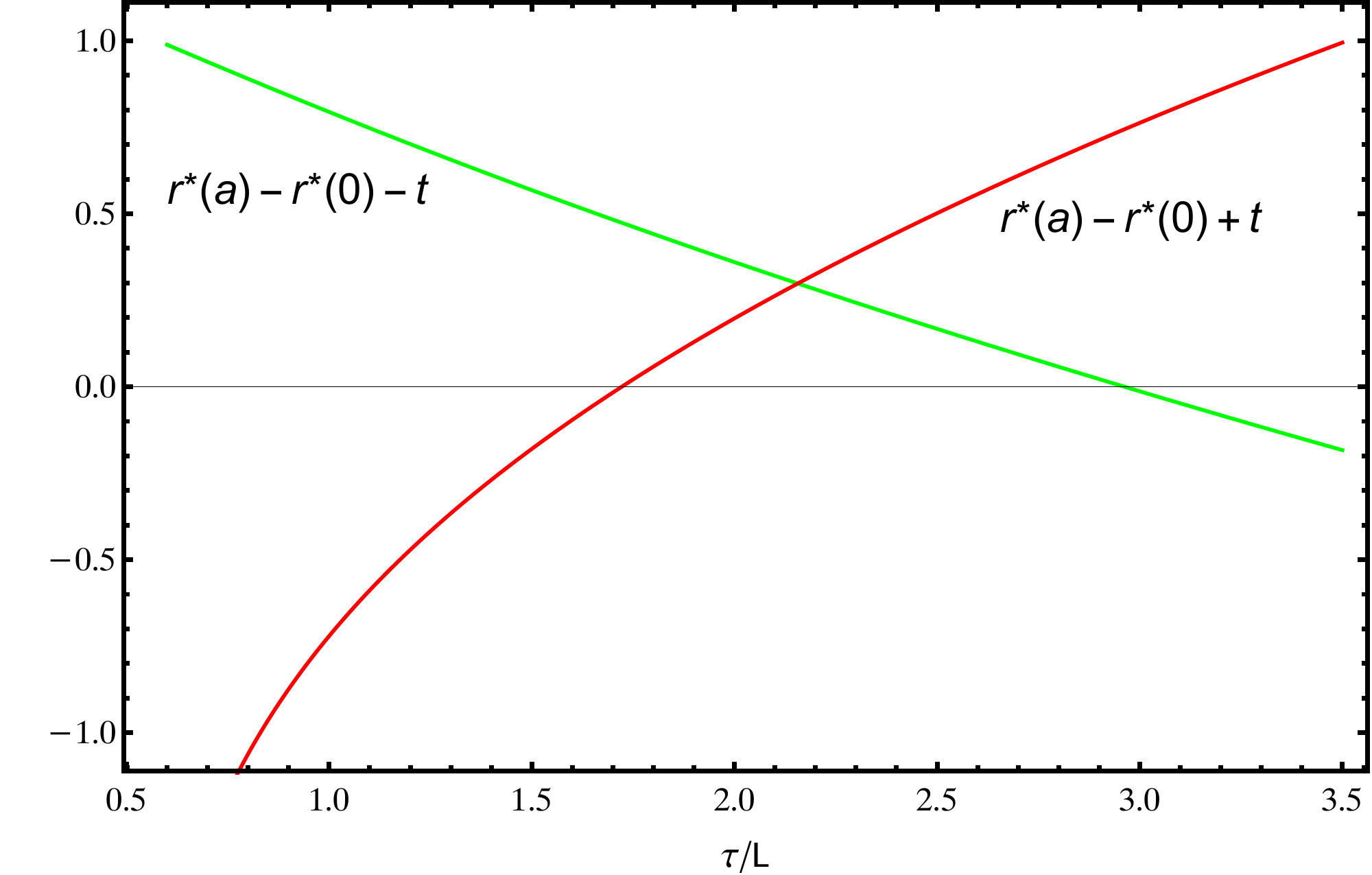}\\
	\caption{The green line represents the function $r^{*}(a)-r^{*}(0)-t$, and the red line is for $r^{*}(a)-r^{*}(0)+t$. We consider the evolution after the brane crosses the horizon. We see clearly that when $0.5L<\tau<1.73L$, there is one future joint term and we denote the positon to be $r_{m1}$. For $1.73L<\tau<2.97L$, there is no joint term, the WDW patch intersects the two singularities. For $\tau>2.97L$, the WDW patch has a past joint term with its location at $r_{m2}$. }
	\label{penr}
\end{figure}

For the CA case, we first need to determine the WDW patch, which depends on the two functions $r^{*}(a)-r^{*}(0)-t$ and  $r^{*}(a)-r^{*}(0)+t$.  We plot both functions in Fig.~\ref{penr}. The WDW patch forms the past null joint when $r^{*}(a)-r^{*}(0)-t<0$ and the future null joint when $r^{*}(a)-r^{*}(0)+t<0$. From Fig.~\ref{penr}, we can find that the critical time $\tau_{c}$ when the past joint term forms is $\tau_{c}=2.97L$. 

\begin{figure}\label{fdsaf}
	\includegraphics[width=0.5\textwidth]{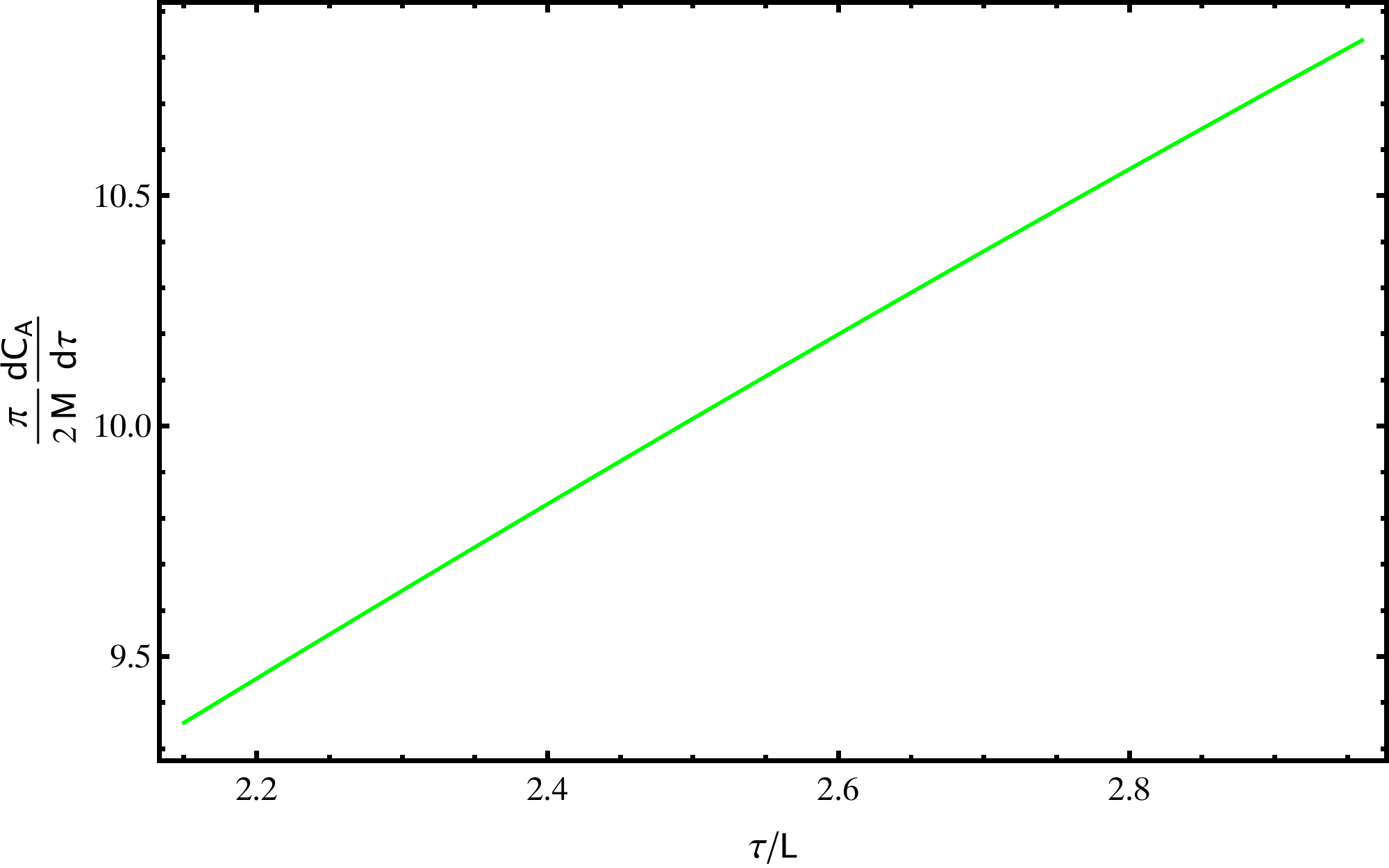}\\
	\includegraphics[width=0.5\textwidth]{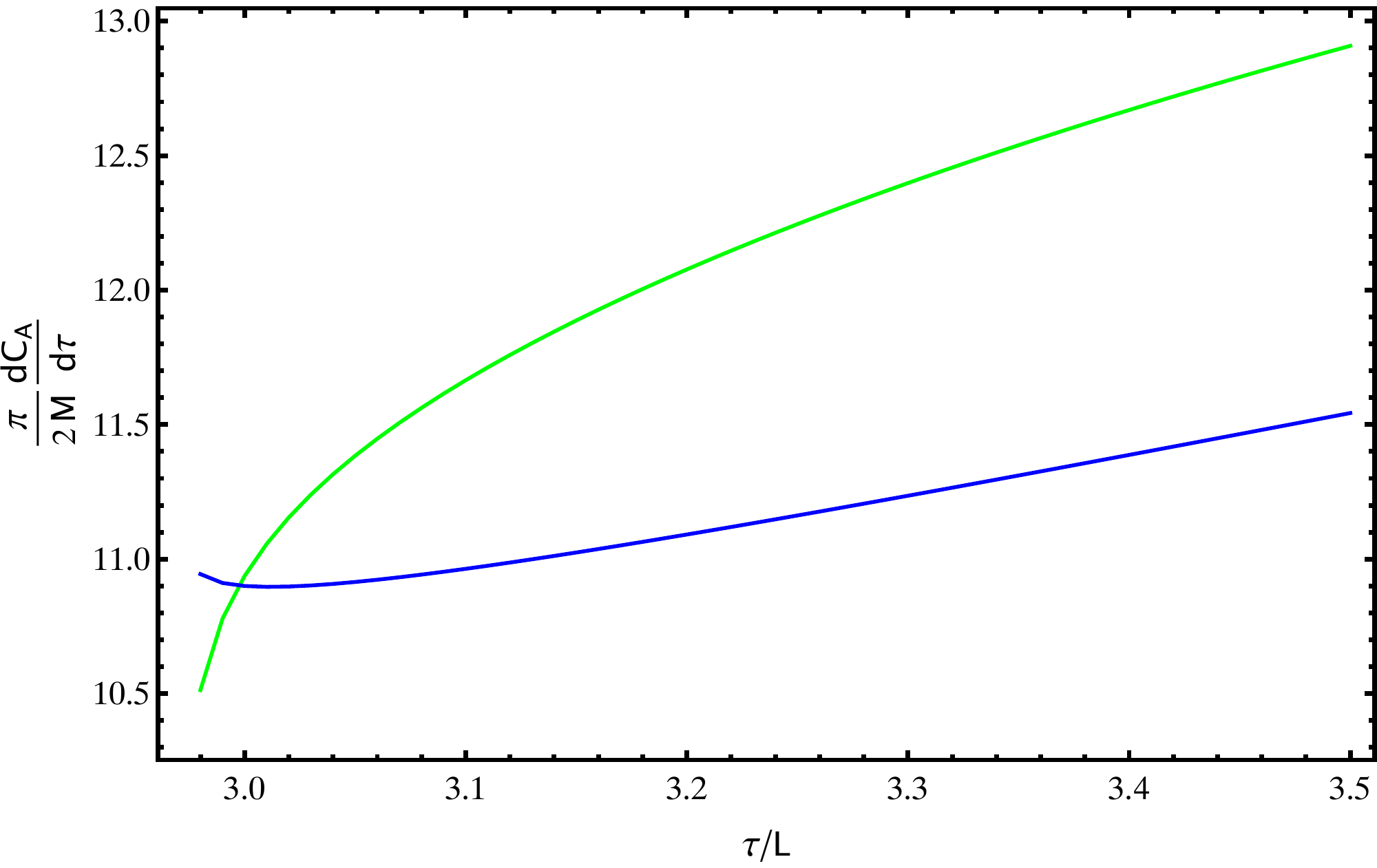}\\
	\caption{The complexity growth rate in the CA conjecture for the flat brane universe. Top panel: complexity growth rate for $\tau<\tau_{c}$. Bottom panel: complexity growth rate for $\tau>\tau_{c}$. The green curve is the total complexity growth, and the blue curve is the complexity growth due to the volume expansion. We see that the contribution comes mainly from the volume expansion part.}
	\label{efsf}
\end{figure}

In order to compare with the static case, we consider the evolution from $t=0$. At $t=0$, we prepare the TFD state $| TFD(0) \rangle_{FRW}$ and then evolve it along the $\tau$ direction. According to Eq.~(\ref{tandtau}), when $t=0$, $\tau$ starts from $\tau=2.16L$. Before the critical time $\tau_c$, the complexity evolution totally comes from the expansion of the volume, which is shown in the top panel of Fig.~\ref{efsf}. After the critical time, as one can see from the bottom panel of Fig.~\ref{efsf}, the main contribution of the growth rate is also from the volume expansion.


\section{Summary of Results and Discussion}

In this paper we have studied the behaviors of the complexity growth rate during some kinds of cosmological evolution in the context of the AdS/CFT correspondence and the brane world framework in an AdS Schwarzschild black hole background. We have considered a closed universe and a flat universe by using then CV and CA conjectures. Here we summarize our analysis and main results.

In section II, we have investigated the complexity growth rate of a TFD state defined on the FLRW type slice located at the asymptotic AdS boundary. Both the CV and CA conjectures give a similar result. We also note that for the CV case, there are also additional sub-leading terms and one particular contribution due to the spatial curvature.  The complexity growth rate consists of two parts. For the first part of Eq.(\ref{fsfss}), which we called ``interaction term", the complexity growth rate is decreasing for an expanding background and vice versa. On the other hand, the second part is proportional to the growth rate of the spatial volume of boundary and is called ``volume term".

The behavior of the ``interaction term" can be understood as follows. For the TFD state, after preparing the $| TFD(0) \rangle$ state by Euclidean path integral at $t=0$, one considers that there are two localized operators $O_{L}(x)$ and $O_{R}(x)$ at left and right boundaries, respectively. As time evolves, because of the interaction in the right boundary system, $O_{R}$ affects and correlates with more and more degrees of freedom in the right side. The original correlation between $O_{L}$ and $O_{R}$ is distributed among many degrees of freedom, and $O_{L}$ correlates with many other operators on the right besides $O_{R}$. So the correlation between $O_{L}(t,x)$ and $O_{R}(t,x)$ decreases, which can be easily seen from the decrease of mutual information and correlation function. Meanwhile, the system becomes more complex because the original $O_{R}$ is scrambling into many degrees of freedom. These explain the complexity growth for the TFD state. 
When the space is expanding or contracting, the spreading of $O_{R}$ will decrease or increase, respectively. So, while the complexity still grows, the growth rate will slow down or increase, depending on the evolution of the background. 

For the volume term of Eq.(\ref{fsfss}), we see that there is some divergence in the complexity growth rate when $R_{max} \to \infty$. To be more concrete, we introduce the UV cutoff $\epsilon=L/R_{max}$, and the volume term can be rewritten as 
\begin{equation}
\frac{1}{d-1}\frac{\mathrm{d}}{\mathrm{d}\tau} (\frac{L^{d-1} a^{d-1}}{\epsilon^{d-1}})\,.
\end{equation}
One finds that such divergent term obeys a volume law. This result is quite natural from the field theory point of view. Note that from the field theory definition of complexity, the leading contribution of complexity is indeed the volume law.  For example, with an appropriate choice of the cost function and the reference frequency, the complexity of free field theory is given by~\cite{Jefferson:2017sdb}
\begin{equation}
\mathcal{C}=\frac{V}{\delta^{d-1}}\,,
\end{equation}
with $\delta$ the UV cutoff and V the volume of the space where the field theory is defined. A simple physical picture is as follows. As the volume expands, there appears many new degrees of freedom which also appear in the computation process. So the complexity will increase and be proportional to the growth of the background volume.

In Section III, we focus on the brane cosmology for which the FLRW universe lives in a brane located at finite radius of an AdS black hole. We have considered both the closed universe $k=1$ and the flat universe $k=0$. Now as the conformal radiation field is coupled to gravity, there are some modifications to the above two terms. It is worthy pointing out that in the brane cosmology setup, the complexity growth rate is free of UV divergence. 
The behavior of the complexity growth using CV duality and CA duality is different. For the CV duality, the main contribution always comes from the ``volume part". For the closed universe with $k=1$, the complexity growth behavior also depends on the horizon radius $r_{h}$. When $r_h/L$ is small, the growth rate first increases and then decreases. When $r_{h}/L$ is large, the growth rate decreases monotonically and is always negative due to the contraction of the volume. 
For the flat case with $k=0$, the universe on the brane expands. It has been found that the complexity growth rate is positive due to the expansion of the volume, and in both cases the ``interaction part" plays little role. But for the CA duality, there is some competition between the ``interaction part" and the ``volume part". For the $k=1$ case, the complexity growth rate is at first negative, but after the critical time $\tau_{c}$, it becomes positive, which is quite different from the CV calculation. For the $k=0$ case, the complexity growth rate always grows with its contribution mainly from the volume expansion.

\begin{figure}
	\includegraphics[width=0.5\textwidth]{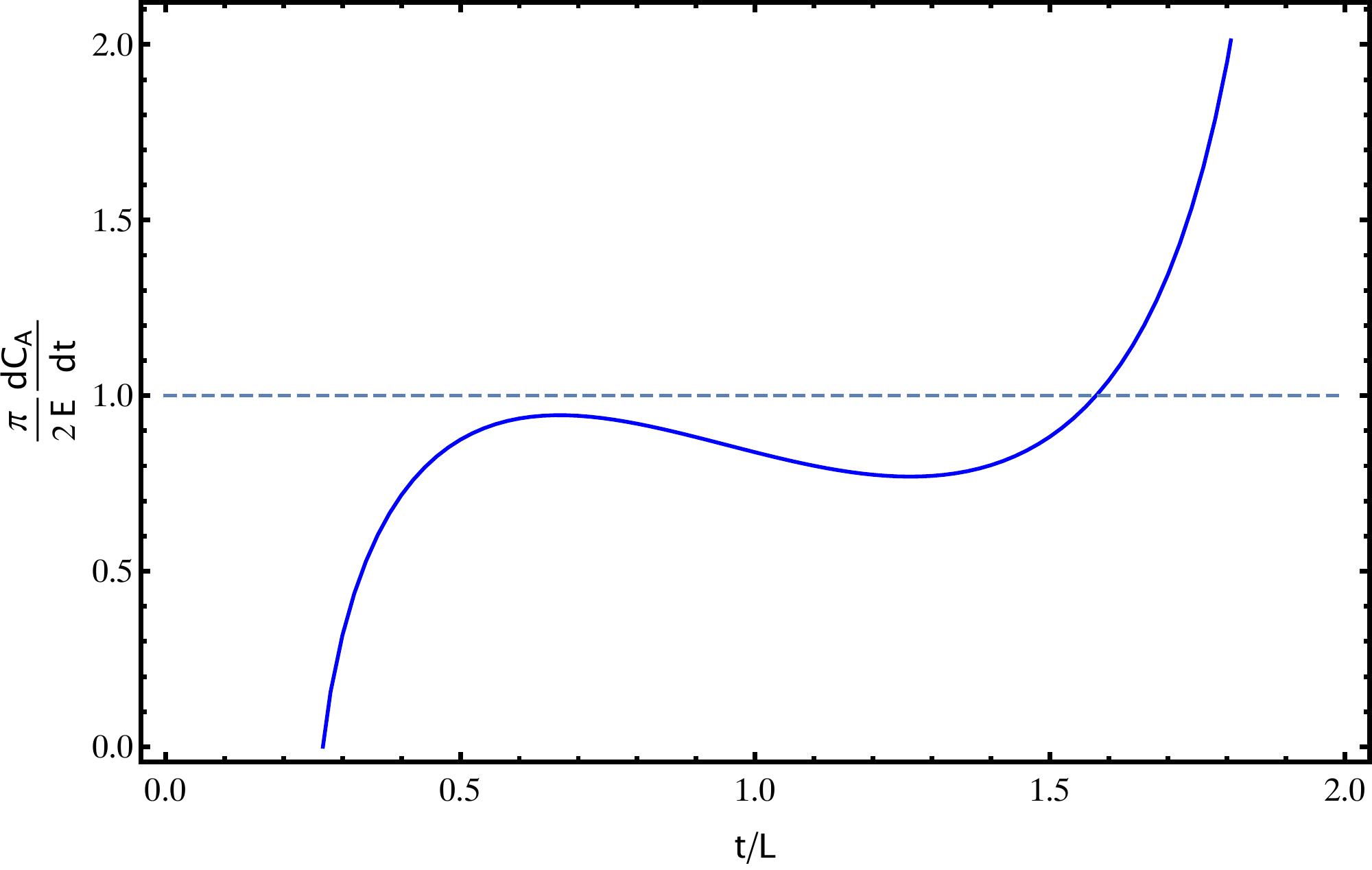}\\
	\caption{The complexity growth rate in the CA conjecture for the closed brane universe with $r_h=L$ and $d=4$. Before time $t=1.6L$, the complexity growth satisfies the Lloyd bound, while after that time as the brane cross the horizon, the complexity growth rate increase quickly again and thus violates the Lloyd bound}
	\label{lloyd}
\end{figure}

In this work, we have denoted the mass of the black hole by $M$. But we should note that $M$ is not the energy of the expanding/contracting universe on the brane. 
So it does not relate to the Lloyd bound. The physical energy of the brane universe was given by the author of Ref.~\cite{Verlinde:2000wg}:
\begin{equation}
E=M \frac{L}{a}\,,
\end{equation}
which depends on the evolution of the universe. The relation between total boundary time $t=\tau_{L}+\tau_{R}=2\tau$ and the complexity growth rate divided by the physical energy, $\frac{1}{2E} \frac{dC_{A}}{dt}$, is presented in Fig.~\ref{lloyd} for $k=1$ case. 
Interpretating complexity growth as computation, there is a physical bound for the growth rate conjectured by Lloyd~\cite{Lloyd:2000}. Previous studies considered Lloyd bound in a static boundary~\cite{Brown:2015bva,Brown:2015lvg, Carmi:2017jqz}. Here we would like to discuss the Lloyd bound when the background is changing over time, more specifically during the cosmic evolution we have studied. The $k=1$ case is of particular interest. The universe is now in a contracting phase, and therefore the contribution from the ``volume part" is negative. So the vast growth of complexity growth rate after the critical time $\tau_{c}$ is due to the interaction among the degrees of freedom. 
%
%
However, as shown in Fig.~\ref{lloyd}, when the brane is near the horizon, the complexity growth rate becomes badly divergent hence the Lloyd bound is violated. As this complexity growth divergence is due to the first ``interaction part" of Eq.(\ref{fsfss}), it becomes very strange why the computation can be so fast, which is far beyond the physical constraint of the energy-time uncertainty relation. In Ref.~\cite{Savonije:2001nd}, the author argued that when the brane crosses the horizon, the Casimir energy due to quantum corrections will no longer be small such that this period may not be trustworthy. In our work, based on the complexity growth rate, we give further evidence that there might be some inconsistency when the brane crosses the horizon.  

%
Many open questions and challenges remain. The authors of Ref.~\cite{Akhavan:2018wla} investigated the complexity growth rate for the $T\bar{T}$ deformed CFT on the boundary at finite radius. They found that in order to make the late time complexity growth rate to satisfy the Lloyd bound, one has to introduce a corresponding cutoff surface inside the horizon with its position determined by the outside cutoff surface.
For $k=0$ case, the position of the brane inside the horizon, say at $r_0$, and the cutoff radius, say at $r_c$, have a simple relation for the $AdS_{5}$ case.
\begin{equation}
r_{0} r_{c}^{2}=r_{h}^{3}\,.
\end{equation}
It means that in the construction of the bulk from the boundary evolution, when the evolution is changed, for example, from $H$ to $H_{T\bar{T}}$, the bulk should be changed accordingly, and an additional brane inside the horizon is formed. In our brane-world scenario, similar thing may happen.
It will be interesting to investigate where the new brane is. The introduction of the inside brane might provide a mechanism to prevent the brane from entering the horizon and to cure the inconsistency we found in Fig.~\ref{lloyd}. 

Note that in this work we have only investigated the complexity growth rate on the expanding/contracting universe from the holographic side. It will be also important to understand our results from the field theory point of view. We could use the Fubini-Study metric~\cite{Chapman:2017rqy} to define the complexity and to check if the complexity of free field theory defined on the expanding/contracting background will exhibit a similar behavior as our holographic results. We leave the analysis to the future work. In the present paper, we have only considered a simple case to obtain the brane cosmology in the Schwarzschild black hole. There are many complicated constructions based on other black holes, such as Refs.~\cite{Cai:2001ur,Cai:2001ja,Xu:2019gzt}. It would also be interesting to study the complexity for a generic FLRW background and to see if there are new features.


Note added: As this paper was in preparation, there has been an related paper studying the holographic complexity in FLRW spacetimes~\cite{Caginalp:2019fyt}. In contrast to our setup, that paper considered a holographic screen in the FLRW universe, and investigated the complexity growth rate of a CFT defined on the screen.

\begin{acknowledgments}
Y.-.S An would like to thank Zhuo-Yu Xian for valuable discussion on the result. L.Li is supported by the Chinese Academy of Sciences (CAS) Hundred-Talent Program.
Y.X. Peng is supported in part by the National
Postdoctoral Program for Innovative Talents with Grant No. BX201700259.
R.-G. Cai was supported in part by the National Natural Science Foundation of China Grants Nos.11435006, 11647601,
 11821505, 11851302, 11847612 and by the Key Research Program of Frontier Sciences of CAS。 
\end{acknowledgments}

\bibliographystyle{utphys}
\bibliography{ref}

\end{document}